\documentclass[aps,prd,twocolumn,showpacs,floatfix,preprintnumbers,amsmath,amssymb,nofootinbib, notitlepage]{revtex4-1}

\input epsf
\usepackage{graphicx}
\usepackage{color}
\usepackage{mathtools}
\usepackage{mathbbol}
\usepackage{longtable}
\usepackage{multirow}

\newcommand{\beq}{\begin{equation}}
\newcommand{\eeq}{\end{equation}}
\newcommand{\barr}{\begin{eqnarray}}
\newcommand{\earr}{\end{eqnarray}}

\newcommand{\rme}{\textrm{e}}

\newcommand{\apjl}{Astrophys. J. Lett.}
\newcommand{\aap}{Astron. Astrophys.}

\newcommand{\mnras}{Mon. Not. R. Astron. Soc.}

\newcommand{\jcap}{J. Cosm. Astropart. Phys.}
\newcommand{\apss}{Astrophys. and Space Science}

\newcommand{\codename}{\texttt{DMDIST}}

\usepackage{color}

\newcommand{\lsim}{\mathrel{\hbox{\rlap{\lower.55ex\hbox{$\sim$}} \kern-.3em \raise.4ex \hbox{$<$}}}}
\newcommand{\gsim}{\mathrel{\hbox{\rlap{\lower.55ex\hbox{$\sim$}} \kern-.3em \raise.4ex \hbox{$>$}}}}

\begin{document}
\title{Testing dark matter interactions with CMB spectral distortions}
\author{Yacine Ali-Ha\"imoud}
\affiliation{Center for Cosmology and Particle Physics, Department of Physics,
New York University, New York, NY}
\date{\today}

\begin{abstract}
Possible interactions of dark matter (DM) with Standard Model (SM) particles can be tested with spectral distortions (SDs) of the cosmic microwave background (CMB). In particular, a non-relativistic DM particle that scatters elastically with photons, electrons or nuclei imprints a negative chemical potential $\mu$ to the CMB spectrum. This article revisits the first study of this effect, with an accurate treatment of heat exchange between DM and SM particles. We show that the instantaneous-decoupling approximation made in the original study systematically and significantly underestimates the amplitude of SDs. As a consequence, we derive tighter upper bounds to the DM-SM elastic-scattering cross section for DM masses $m_\chi \lesssim 0.1$ MeV, from the non-detection of $\mu$-distortions by FIRAS. We also show that a future instrument like PIXIE, sensitive to $|\mu| \sim 10^{-8}$, would be able to probe DM-SM cross sections much smaller than first forecasted, and orders of magnitude below current upper limits from CMB-anisotropy data, up to DM masses of $\sim 1$ GeV. Lastly, we study the sensitivity of SDs to the electric and magnetic dipole moments of the DM. Although SDs can place non-trivial constraints on these models, we find that even future SD experiments are unlikely to improve upon the best current bounds. This article is accompanied by the public code \codename, which allows one to compute CMB SDs for generic particle-DM models, specified by their cross sections for elastic scattering with and annihilation into SM particles.
\end{abstract}

\maketitle

\section{Introduction}

The cosmic microwave background (CMB) is a sensitive calorimeter: heating or cooling of the photon-baryon plasma at $z \lesssim 2 \times 10^6$ would distort its frequency spectrum away from a perfect blackbody \cite{ZS_69, SZ_69, SZ_70, SZ_70b}. Several processes lead to guaranteed spectral distortions (SDs) in the standard cosmological scenario \cite{Chluba_16}. First, Thomson (and Coulomb) scattering lead to a systematic heat flow from photons to non-relativistic electron-baryons, and as a consequence, to a negative chemical potential $\mu$ and Compton-$y$ distortion \cite{Chluba_12}. Second, free-bound and line radiation emitted in the process of cosmological recombination imprints features on the CMB spectrum \cite{Rubino_08, Sunyaev_09}. Both of these distortions are of order a few times the baryon-to-photon ratio $n_b/n_\gamma \sim 10^{-9}$ and can be computed very accurately \cite{YAH_13, Chluba_16b}. Third, the dissipation of acoustic waves resulting from photon diffusion \cite{Silk_68} generates a positive chemical potential $\mu$ and Compton $y$-parameter \cite{Barrow_91, Daly_91, Hu_94, Chluba_12, Pajer_13}. The amplitude of this guaranteed distortion depends on the power of primordial perturbations on comoving scales $1 ~ \textrm{Mpc}^{-1} \lesssim k \lesssim 10^4~ \textrm{Mpc}^{-1}$, which are not easily accessible by other observations. Last but not least, at $z \lesssim 10$, CMB photons get up-scattered in the hot, re-ionized intergalactic medium, leading to a Compton-$y$ distortion of order $\sim 10^{-6}$ \cite{Hill_15}. All of these distortions are below the current upper limits from FIRAS, of order $\sim 10^{-5}-10^{-4}$ \cite{Fixen_96}.

In addition to these guaranteed signals, beyond-the-standard-model scenarios could produce additional SDs. For instance, a dark matter (DM) particle annihilating or decaying to electromagnetically-interacting particles could distort the CMB spectrum \cite{McDonald_00}. In this paper, we focus on the effect of a non-relativistic dark matter (DM) particle elastically scattering with photons, electrons or nuclei. Just like electron-baryons do through Thomson scattering, such a particle would systematically extract heat from the photon-baryon plasma. The amplitude of the distortion can be as large as the DM-to-photon number ratio $n_\chi/n_\gamma$, inversely proportional to the DM mass $m_\chi$. This effect, first pointed out and studied in Ref.~\citep{ACK15} (hereafter ACK15), can therefore be used to probe light (sub-GeV, or even sub-MeV) DM particles, which are, for the time being, out of reach of most direct-detection experiments.

In ACK15, we derived upper limits to the cross section of DM with Standard Model (SM) particles given the non-detection of SDs by FIRAS \cite{Fixen_96}, and forecasted the reach of a future SD experiment like PIXIE \cite{Kogut_11}. The main ingredient in these limits and forecasts is the heat-exchange rate between DM and the photon-baryon plasma. This rate depends on the velocity distribution of DM particles as they thermally decouple from the plasma. Lacking a detailed description of this distribution, we had simply approximated thermal decoupling as occurring instantaneously at a characteristic decoupling redshift. This approximation moreover allowed us to derive simple analytic expressions for the detectable scattering cross section for a given sensitivity to SDs. 

The velocity distribution of DM depends not only on its interactions with SM particles, but also on its self-interactions. Indeed, frequent self-scattering leads to a re-shuffling of DM velocities towards the maximum-entropy, thermal Maxwell-Boltzmann (MB) distribution. If the timescale for velocity re-distribution through self-interactions is longer than the expansion timescale, the DM velocity distribution need not be MB after the DM thermally decouples from SM particles. In a recent work, we studied the evolution of the background DM velocity distribution, by solving the Boltzmann equation with a Fokker-Planck collision operator \cite{YAH_19}. Within this approximation for the collision operator, the background (homogeneous and isotropic) DM velocity distribution, relevant to CMB SDs, is exactly MB if DM scatters with photons only \cite{Bertschinger_06}. For a DM particle scattering with baryons, we found in Ref.~\cite{YAH_19} that the heat-exchange rate can differ by up to a factor $\sim 2-3$ between the limiting cases of negligible self-interactions and strong self-interactions. The largest difference occurs for DM-baryon elastic cross sections with a strong velocity dependence, and small DM-to-baryon mass ratio $m_\chi/m_b \ll 1$. For cross sections with shallower velocity dependence and/or mass ratios $m_\chi/m_b \gtrsim 1$, we found that the MB approximation allows us to compute the heat-exchange rate within a few tens of percent accuracy\footnote{Let us emphasize that these findings only apply to the \emph{background} DM distribution: even in the case of DM-photon scattering, the \emph{perturbed} DM distribution departs from the MB distribution, and these departures could significantly impact CMB-anisotropy limits to DM-baryon and DM-photon scattering.}. 

Armed with this deeper understanding of the DM velocity distribution and its impact on heat exchange, in this paper we revisit the calculation of ACK15. We evaluate the heat-exchange rate numerically, assuming the DM has a thermal MB distribution, with the understanding that in some cases the results are only accurate within factors of order unity. We show that the instantaneous-decoupling approximation made in ACK15 \emph{significantly} underestimates the net amount of photon-baryon cooling during the $\mu$-era, as a result of the residual heat exchange persisting after the characteristic decoupling time. 
As a consequence, we find that the constraints and especially forecasts of ACK15 are too conservative: we show that a future SD experiment could in fact probe DM-SM interactions orders of magnitude weaker than first estimated in ACK15. 

In addition to revisiting and solidifying the analysis of ACK15, the main goal of this paper is to introduce the public code \codename, dedicated to computing the SDs resulting from DM interactions with SM particles. This code should provide a useful tool to derive limits and make forecasts for specific DM candidates. As an illustration, we compute the FIRAS limits on the DM elastic-scattering cross sections with indvidual scatterers (photons, electrons, and nuclei), with different velocity and energy dependences, and moreover forecast the reach of future SD experiments. As a more realistic example, we also compute the SD resulting from a DM particle with an electric or magnetic dipole moment, accounting for both elastic scattering with plasma particles, as well as annihilations into photons and fermion-antifermion pairs. 

The remainder of this article is organized as follows. After summarizing the basic underlying physics in Sec.~\ref{sec:physics}, we describe our numerical integration methods and illustrate the inaccuracy of the instantaneous-decoupling approximation in Sec.~\ref{sec:numerics}. We derive upper limits and make sensitivity forecasts for the DM scattering cross section with individual particles in Section \ref{sec:limits-PL}, and study the sensitivity of SD experiments to the DM electric and magnetic dipole moments in Sec.~\ref{sec:dipole}. Finally, we discuss the limitations and extensions of this work and conclude in Section \ref{sec:conclusion}. For all numerical applications, we adopt the best-fit cosmological parameters consistent with the latest \emph{Planck} results \cite{Planck_18}.

\section{Basic physics and phenomenology} \label{sec:physics}

\subsection{CMB spectral distortions from heat injection}

Let us first briefly summarize the physics of CMB SDs; for a pedagogical introduction, see e.g.~Ref.~\cite{Chluba_18} and references therein. Suppose some process(es) inject heat into (or extract heat from) the photon-baryon plasma. If heat injection takes place at redshift $ z \lesssim z_\mu \equiv 2 \times 10^6$, bremsstrahlung and double-Compton scattering are unable to efficiently change photon number, and the CMB  spectrum is distorted away from a perfect blackbody \cite{ZS_69, SZ_69, SZ_70, SZ_70b, Chluba_12}. If heat injection is moreover limited to $z \gtrsim z_y \equiv 5.8 \times 10^4$ \cite{Chluba_13}, Compton scattering efficiently redistributes photon energies, so the CMB spectrum retains the Bose-Einstein shape, but acquires a non-zero chemical potential $\mu$. For energy injection at $z \lesssim z_y$, the shape of the distortion depends on the detailed channel(s) of heat injection. If heat is injected into electrons and/or baryons, it is transferred to photons through Coulomb and Thomson scattering, and the resulting distortion takes on a universal shape, that of a Compton-$y$ distortion \cite{SZ_69}. If heat is injected through different channels, e.g.~through a direct photon injection \cite{Chluba_15}, or through non-Thomson elastic scattering with a non-relativistic particle, the shape of the late-time distortion is not the Compton-$y$ distortion. Note that if energy injection occurs at intermediate epochs $z \sim z_y$, the distortion is neither pure $\mu$ nor pure $y$, and must be computed by solving the Boltzmann equation numerically \cite{Chluba_12}.

In this paper, we focus exclusively on $\mu$-distortions, thus on energy injection/extraction at $z_y \lesssim z \lesssim z_\mu$. For heat injection with a net volumetric rate $\dot{\mathcal{Q}}$, the amplitude of the chemical potential is approximately \cite{Chluba_13}:
\barr
\mu &\approx& \int \frac{da}{a} \frac{\dot{\mathcal{Q}}}{H \rho_\gamma} \mathcal{G}_\mu(a) , \label{eq:mu}\\
\mathcal{G}_\mu(a) &\equiv & 1.4 \left( 1- \rme^{ - (a z_y)^{-1.88}} \right) \rme^{- (a z_\mu)^{-5/2}}, \label{eq:Green-mu}
\earr
where $a = 1/(1+z)$ is the scale factor, $H(a)$ is the Hubble rate, and $\rho_\gamma(a)$ is the energy density of CMB photons. The Green's function $\mathcal{G}_{\mu}(a)$ given in Eq.~\eqref{eq:Green-mu} is approximate; see \cite{Chluba_14} for more accurate calculations.

\subsection{Cooling from DM scattering}

\subsubsection{General considerations}

Suppose a fraction $f_\chi$ of the DM scatters with either photons or electrons/baryons. As pointed out in ACK15, as long as the particle $\chi$ is non-relativistic, this interaction generates a heat exchange between DM and photons, similar to the standard cooling effect caused by baryons themselves \cite{Chluba_12}. This can in principle be observable through CMB spectral distortions. 

For simplicity, we  approximate the DM velocity distribution with a Maxwell-Boltzmann (MB) distribution at temperature $T_\chi$. This assumption holds either as long as the DM is thermally coupled to the photon-baryon plasma, or is efficiently self-interacting. If neither criterion is satisfied, the MB approximation may lead to order-unity inaccuracies in the heat-exchange rate when the DM is lighter than the scatterer \cite{YAH_19}, and our results should thus be interpreted as accurate within a factor of order unity. 

Since we restrict ourselves to the $\mu$-era, we may neglect the peculiar velocity of DM with respect to the photon-baryon plasma, as it is small relative to the baryon thermal velocity at $z \gtrsim 10^4$ \cite{Dvorkin_14}. We limit ourselves to the non-relativistic limit, $T_\chi/m_\chi \ll 1$. For this limit to hold at $z \approx 2 \times 10^6$, at which point $T_\gamma \approx 0.5$ keV,  we restrict ourselves to $m_\chi >$ 1 keV.

We denote by $\dot{\mathcal{Q}}_\chi$ the rate of DM heating per unit volume:
\beq
\dot{\mathcal{Q}}_\chi \equiv \frac32 n_\chi a^{-2} \frac{d}{dt}(a^2 T_\chi). \label{eq:dTchi/da}
\eeq
If this heating is generated by direct elastic scattering with photons, then the photon volumetric heating rate is just $\dot{\mathcal{Q}} = - \dot{\mathcal{Q}}_\chi$. If DM is heated by scattering with electrons/baryons, this relation still holds, as long as they are tightly thermally coupled to photons, i.e.~for $z \gtrsim 200$. Note that this equality neglects the dissipation of peculiar velocities into heat \cite{Munoz_15}.

We define the dimensionless heat-exchange rate 
\beq
Q \equiv \frac{\dot{\mathcal{Q}}_{\chi}}{\frac32 n_\chi H T_\gamma} = \frac{T_\chi}{T_\gamma}\frac{ d \ln(a^2 T_\chi)}{d \ln a}. \label{eq:Q-def}
\eeq
This quantity is always positive (provided the DM does not start hotter than the plasma), and is such that $Q \rightarrow 1$ when the DM is tightly coupled to the photon-baryon plasma, and $Q < 1$ otherwise. In terms of $Q$, the integrand in Eq.~\eqref{eq:mu} is 
\beq
\frac{\dot{\mathcal{Q}}}{H \rho_\gamma} = - \frac32 f_\chi \frac{\rho_c T_\gamma}{\rho_\gamma m_\chi} Q \approx - 1.7 \times 10^{-6} f_\chi \frac{\textrm{MeV}}{m_\chi} Q.
\eeq
Note that the last equality assumes that the photon temperature is unperturbed from its standard evolution, $T_\gamma = T_0/a$, an assumption we justify further in Sec.~\ref{sec:pheno}.

The maximum $\mu$ distortion produced by elastically scattering DM is obtained by setting $Q = 1$ at all times in Eqs.~\eqref{eq:mu}. Numerically, we find
\barr
|\mu_{\max}| \approx 8.6 \times 10^{-6} ~ f_\chi \frac{\textrm{MeV}}{m_\chi}. \label{eq:mu-max}
\earr
The 95\%-confidence upper limit to spectral-distortion derived from FIRAS observations is $|\mu| < 9 \times 10^{-5}$ \cite{Fixen_96}. We hence see that FIRAS is only sensitive to DM masses below $m_{\max} \sim  0.1~f_\chi$ MeV, as lighter particles would not sufficiently distort the CMB, even if they were tightly thermally coupled throughout the $\mu$-era. On the other hand, a sensitivity to distortions at the level of $|\mu| \sim 10^{-8}$, as could be achieved by proposed instruments like PIXIE \cite{Kogut_11}, would allow to probe scattering DM with mass up to $m_{\max} \sim f_\chi$ GeV \cite{ACK15}.

\subsubsection{DM scattering with photons}

Consider a non-relativistic DM particle elastically scattering with photons, with a momentum-transfer cross section dependent on photon energy $\sigma_{\chi \gamma}(E_\gamma)$. 
The simplest route to obtain the net heating rate is to first compute it at zero DM temperature, accounting for stimulated scatterings, and find the corresponding expression for finite DM temperature (to lowest order in $T_\chi/m_\chi$) using detailed balance. Doing so, we obtain the following DM volumetric heating rate, in natural units:
\barr
\dot{\mathcal{Q}}_\chi  &=& \frac32 n_\chi \Gamma_{\chi \gamma} (T_\gamma - T_\chi),\\
\Gamma_{\chi \gamma}  &\equiv& \frac83 \frac{\rho_\gamma}{m_\chi} \langle \sigma_{\chi \gamma} \rangle , \label{eq:Q-photon}\\
\langle \sigma_{\chi \gamma} \rangle &\equiv& \frac{15}{4 \pi^4} \int_0^{\infty} \frac{d\epsilon}{\rme^\epsilon -1} \frac{d}{d\epsilon} \left[\epsilon^4 \sigma_{\chi \gamma}(\epsilon T_\gamma) \right]. \label{eq:sigma_gamma-av}
\earr
This expression approximates the CMB spectrum as a perfect blackbody at temperature $T_\gamma$, thus neglecting the feedback of small SDs into the heat-exchange rate. 

Following ACK15, we specialize to cross sections with a power-law dependence on photon energy. We adopt the following normalization convention (different from that of ACK15):
\beq
\sigma_{\chi \gamma}(E_\gamma) = \sigma_* \left(\frac{E_\gamma}{m_\chi}\right)^n, \label{eq:sigma-photon}
\eeq
where $n$ is an even positive integer. In that case, Eq.~\eqref{eq:sigma_gamma-av} can be computed explicitly:
\barr
\langle \sigma_{\chi \gamma} \rangle = \frac{15}{4 \pi^4} (4 + n)! ~\zeta(4 + n) \sigma_* \left(\frac{T_\gamma}{m_\chi}\right)^n, 
\earr
where $\zeta$ is the Riemann zeta funtion. For a Thomson-like cross section $\sigma_{\chi \gamma} = \sigma_0$, this expression reduces to $\langle \sigma_{\chi \gamma} \rangle = \sigma_0$, and the net heating rate \eqref{eq:Q-photon} reduces to the well-known expression for Compton heating \cite{Weymann_65}.

\subsubsection{DM scattering with non-relativistic nuclei or electrons}

Consider DM particles scattering off non-relativistic scatterers $s$ (either nuclei or electrons) of mass $m_s$, temperature $T_s$, abundance $n_s$ and mass density $\rho_s = m_s n_s$. For a general velocity-dependent momentum-transfer cross section $\sigma_{\chi s}(v)$, where $v$ is the DM-scatterer relative velocity, the heat exchange rate is \cite{Dvorkin_14, Boddy_18, YAH_19}
\barr
\dot{\mathcal{Q}}_\chi &=& \frac32 n_\chi \Gamma_{\chi s} (T_s - T_\chi),\\
\Gamma_{\chi s}& \equiv& \frac23 \frac{m_\chi \rho_s}{(m_\chi + m_s)^2} \frac{\langle \sigma_{\chi s}(v) v^3 \rangle}{v_{\rm th}^2}  ,    \label{eq:Qbaryon} \\
\langle \sigma(v) v^3 \rangle&\equiv& \sqrt{\frac{2}{\pi}} v_{\rm th}^{-3} \int dv ~ v^5 ~ \sigma_{\chi s}(v)\exp\left[-\frac{v^2}{2 v_{\rm th}^2}\right],\\
v_{\rm th}^2 &\equiv& T_\chi/m_\chi + T_s/m_s.
\earr
Note that we are neglecting DM-baryon bulk relative velocities relative to the thermal relative motions, which is accurate for $z \gtrsim$ few times $10^4$ \cite{Dvorkin_14}.

We specifically consider velocity-dependent momentum-transfer cross sections of the form $\sigma_{\chi b}(v) = \sigma_* v^n$, where $n$ is an even integer. In that case, for $n \geq -4$, we have \cite{Dvorkin_14, Boddy_18, YAH_19}
\barr
\frac{\langle \sigma_{\chi s}(v) v^3 \rangle}{v_{\rm th}^2} = \frac{2^{\frac{5 +n}2}}{\sqrt{\pi}}(2 + n/2)! ~ \sigma_* v_{\rm th}^{n+1}. \label{eq:sig-v-powlaw}
\earr
Note that such power-law cross sections are typically only valid in the perturbative limit, and that non-perturbative effects can lead to more complex dependences, due to resonances and antiresonances \cite{Xu_21}.

\subsection{DM scattering phenomenology}\label{sec:pheno}

The DM temperature satisfies the ordinary differential equation (ODE)
\beq
a^{-2} \frac{d}{dt} (a^2 T_\chi) =  \sum_{s = \gamma, e, p, \rm {He}} \Gamma_{\chi s}(T_s - T_\chi).
\eeq
In principle this ODE should be solved alongside the corresponding equations for $T_\gamma, T_e, T_p, T_{\rm He}$. In practice, as long as $n_\chi/n_\gamma \ll 1$, the photon temperature deviates very little from its unperturbed evolution, $T_\gamma = T_0/a$, where $T_0 \approx 2.73$ K is the CMB temperature today \cite{Fixen_96}. Indeed, deviations from this evolution are comparable to the amplitude of spectral distortions, which are constrained to be small. Moreover, electrons and nuclei remain tightly thermally coupled together due to Coulomb interactions, so $T_p = T_{\rm He} = T_e$. Lastly, electrons and nuclei are tightly coupled to CMB photons through Compton heating at the redshifts of interest, so that one may assume $T_e = T_\gamma$. In practice, we may therefore assume that the photon-baryon plasma is in quilibrium at temperature $T_\gamma = T_0/a$, and only need to solve for the DM temperature evolution, which satisfies
\barr
a^{-2} \frac{d}{dt} (a^2 T_\chi) &=& \Gamma_{\rm tot} (T_\gamma - T_\chi),  \label{eq:Tchi-ODE}\\
\Gamma_{\rm tot} &\equiv&  \sum_{s = \gamma, e, p, \rm {He}} \Gamma_{\chi s}.
\earr
During radiation domination $\Gamma_{\chi \gamma}/H  \propto a^{-(p+2)}$. Thus, for DM-photon scattering with $p \geq 0$, the DM is initially in thermal equilibrium with photons, and eventually decouples when $\Gamma_{\chi \gamma}/H$ falls below unity. 

For DM scattering with electrons or nuclei, the coefficient $\Gamma_{\chi s}$ itself depends on $T_\chi$. If we assume that either $T_\chi \approx T_\gamma$ or $T_\chi/m_\chi \ll T_\gamma/m_s$ initially, we find, during radiation domination, $\Gamma_{\chi s}/H \propto a^{- (n + 3)/2}$. For $n \geq -2$, this ratio is decreasing with time, implying that the DM is initially thermally coupled to baryons (i.e.~with $T_\chi = T_\gamma$). For $n = -4$, this ratio increases with time, implying that the DM starts thermally decoupled (i.e.~with $T_\chi/m_\chi \ll T_\gamma/m_s$), and eventually couples to baryons when this ratio reaches order unity. We do not consider the latter case in this paper, as it would be best constrained by Compton-$y$ distortions, the accurate computation of which would require accounting for the contributions of bulk relative motions.

All the cases we consider thus share the same general phenomenology. At early times, $z \gtrsim z_{\rm dec}$, where $z_{\rm dec}$ is such that $\Gamma_{\rm tot}/H \big{|}_{z_{\rm dec}} = 1$, the DM is in thermal equilibrium with the photon-baryon plasma, $T_\chi \approx T_\gamma \propto 1/a$, and the dimensionless cooling rate is saturated, $Q \approx 1$. After thermal decoupling, the DM temperature starts decreasing faster, tending towards the adiabatic evolution $T_\chi \propto 1/a^2$. The dimensionless heating rate $Q$ then falls below unity, with a decay rate depending on the dominant interaction. 

\subsection{Approximate treatment for thermal decoupling before the non-relativistic transition} \label{sec:relativistic}

As we will see in the next section, for light enough DM particles and weak enough interactions, a significant SD can be produced even if the DM thermally decouples before the $\mu$ era, i.e.~if $z_{\rm dec} > z_\mu$. We restrict ourselves to DM particles with mass $m_\chi \gtrsim$ keV, which always become non-relativistic at a redshift $z_{\rm nr} > z_\mu$. However, sufficiently light and weakly interacting DM particles may thermally decouple before becoming non-relativistic, i.e.~be such that $z_{\rm dec} > z_{\rm nr} > z_\mu$. If that is the case, their temperature scales as $T_\chi \propto 1/a$ between $z_{\rm dec}$ and $z_{\rm nr}$, and thus remains close to the photon temperature until $z_{\rm nr}$, after which it starts decaying as $1/a^2$ at $z \lesssim z_{\rm nr}$. In order to approximately account for this evolution, we only start solving the temperature ODE \eqref{eq:Tchi-ODE} at $z_{\rm nr}$ (defined such that $T_\gamma/m_\chi|_{z_{\rm nr}} = 1$), starting with initial condition $T_\chi = T_\gamma$. 

Particles that are so weakly coupled that $z_{\rm dec} \gg z_{\rm nr}> z_\mu$ already have a temperature $T_\chi \ll T_\gamma$ by the beginning of the $\mu$ era. As a consequence, during the $\mu$-era the heating rate is $\dot{\mathcal{Q}}_\chi \approx \frac32 n_\chi T_\gamma \Gamma_{\rm tot}$. For interactions with nuclei or electrons, the rate $\Gamma_{\chi s}$ is an increasing function of $T_\chi$ if $n > -2$, and is a decreasing function of $T_\chi$ for $n = -2$, as can be seen from Eq.~\eqref{eq:sig-v-powlaw}. For light enough DM particles decoupling before becoming non-relativistic, their temperature is larger than it would be if naively assuming the standard non-relativistic evolution, as it scales as $1/a$ until the particle becomes non-relativistic. Therefore, for $n > -2$ ($n = - 2$), correctly --  even if approximately -- accounting for the DM temperature evolution during $z_{\rm dec} \leq z \leq z_{\rm nr}$ implies a larger (lower) heat-exchange rate than one would naively obtain if simply extrapolating the non-relativistic behavior. This translates to an increased (decreased) sensitivity to $\sigma_*$ for $n > -2$ ($n = -2$) at low masses, relative to the extrapolation of the $\sigma_*(m_\chi)$ relation at higher masses. We shall indeed see this behavior in subsequent figures, for DM-nuclei and DM-electron scattering. For DM-photon scattering, the interaction rate $\Gamma_{\chi \gamma}$ does not depend explicitly on DM temperature, and there is no equivalent feature corresponding to the transition from $z_{\rm dec} < z_{\rm nr}$ to $z_{\rm dec} > z_{\rm nr}$.

One should keep in mind that our treatment of the relativistic transition is only approximate, and we defer a rigorous analysis to future work.

\section{Numerical evaluation} \label{sec:numerics}

\subsection{ODE integration}

In all cases we consider, the DM starts thermally coupled to the photon-baryon plasma. As long as $\Gamma_{\rm tot}/H \geq 10^5$, we use the tight-coupling approximation,
\beq
T_\chi - T_\gamma \approx \frac{H T_\gamma}{ \Gamma_{\rm tot}}, \ \ \  \  \ \ \ \dot{\mathcal{Q}}_\chi \approx  \frac32 n_\chi H T_\gamma.
\eeq
We start explicitly integrating Eq.~\eqref{eq:Tchi-ODE} at $z = \min(z_{\rm nr}, z_5)$, where $z_{\rm nr} \equiv m_\chi/T_0$ is the redshift at which the DM becomes non-relativistic and $z_5$ is defined through $\Gamma_{\rm tot}/H|_{z_5} = 10^5$. After that time, we solve for $T_\chi/T_\gamma$ using a second-order implicit integrator, and a logarithmic step size in scale factor $d \ln a = 10^{-2}$. We then obtain $\dot{\mathcal{Q}}_\chi = (3/2) n_\chi/ \Gamma_{\rm tot}(T_\gamma - T_\chi)$. We checked the accuracy of our numerical ODE integrator by comparing its output to analytic solutions for DM-photon scattering, which we provide in Appendix \ref{app:analytic}. We find that our numerical solution for $\dot{\mathcal{Q}}_\chi$ is accurate to better than $0.1\%$ across all scale factors. Lastly, we compute the $\mu$-distortion by evaluating the integral \eqref{eq:mu} with a simple trapezoidal rule.

\subsection{Inaccuracy of the instantaneous-decoupling approximation}

In our first study of SDs resulting from elastically scattering DM in ACK15, we made a simple approximation for the dimensionless heating rate $Q$ defined in Eq.~\eqref{eq:Q-def}: we approximated it by a step function, equal to unity for $a< a_{\rm dec}$ and vanishing at $a > a_{\rm dec}$, where $a_{\rm dec}$ is a characteristic thermal decoupling scale factor. This allowed us to provide simple analytic expressions to evaluate the sensitivity of a given SD experiment to DM interactions. While this may a priori seem like a reasonable approximation at the factor-of-a-few level, it turns out that it typically \emph{significantly} under estimates the net heat injection, thus the resulting $\mu$ distortion, as we show below. For definiteness, we shall define $a_{\rm dec}$ as the scale factor at which $\Gamma_{\chi, \rm tot}/H = 1$, where $\Gamma_{\rm tot}$ is evaluated at $T_\chi = T_\gamma$. Note that this definition is slightly different from that of ACK15, but this does not affect any of the conclusions. 

\begin{figure}
\includegraphics[width = \columnwidth]{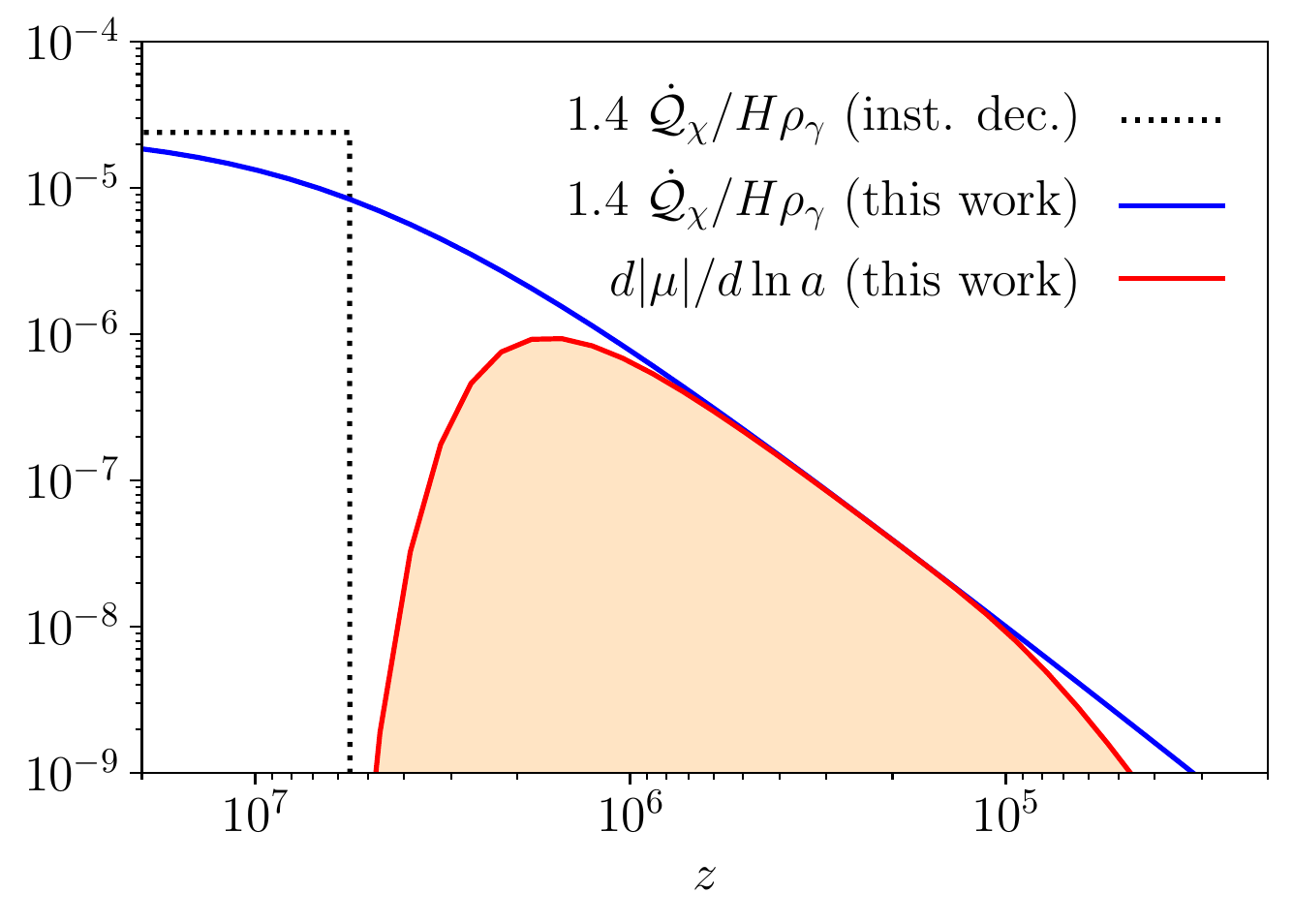}
\caption{Illustration of the inaccuracy of the instantaneous decoupling approximation, for a velocity-independent DM-proton cross section $\sigma_{\chi p} = 1.5 \times 10^{-26}$ cm$^2$, and DM mass $m_\chi = 0.1$ MeV.  The dimensionless quantity $1.4 \dot{\mathcal{Q}}_\chi/H \rho_\gamma$, multiplied by the $\mu$-distortion Green's function, gives the contribution to the chemical potential per logarithmic redshift interval. With the adopted parameters, thermal decoupling occurs at $z_{\rm dec} \approx 5 \times 10^6$, well before the beginning of the $\mu$ era. Therefore, one would conclude that the $\mu$ distortion is completely negligible in the instantaneous-decoupling approximation (dashed line). When correctly computing the heat exchange rate (solid blue line), the residual heat exchange after thermal decoupling leads to a total distortion $|\mu| \approx  10^{-6}$, which is the area under the red curve.}\label{fig:inst_dec}
\end{figure}

In Fig.~\ref{fig:inst_dec}, we show the dimensionless quantity $1.4 \dot{\mathcal{Q}}_\chi/H \rho_\gamma$ computed exactly and in the instantaneous-decoupling approximation, for $m_\chi = 0.1$ MeV and a velocity-independent DM-proton cross section $\sigma_{\chi p} = 1.5 \times 10^{-26}$ cm$^2$. Even though decoupling occurs before the beginning of the $\mu$ era at $z_{\rm dec} \approx 5 \times 10^6$, the residual heat exchange at $z <  z_{\rm dec}$ leads to an appreciable distortion $|\mu| \approx 10^{-6}$. This residual heat exchange is not captured in the instantaneous-decoupling approximation, which in this case would lead to a completely negligible SD. 

\begin{figure}
\includegraphics[width = 1\columnwidth]{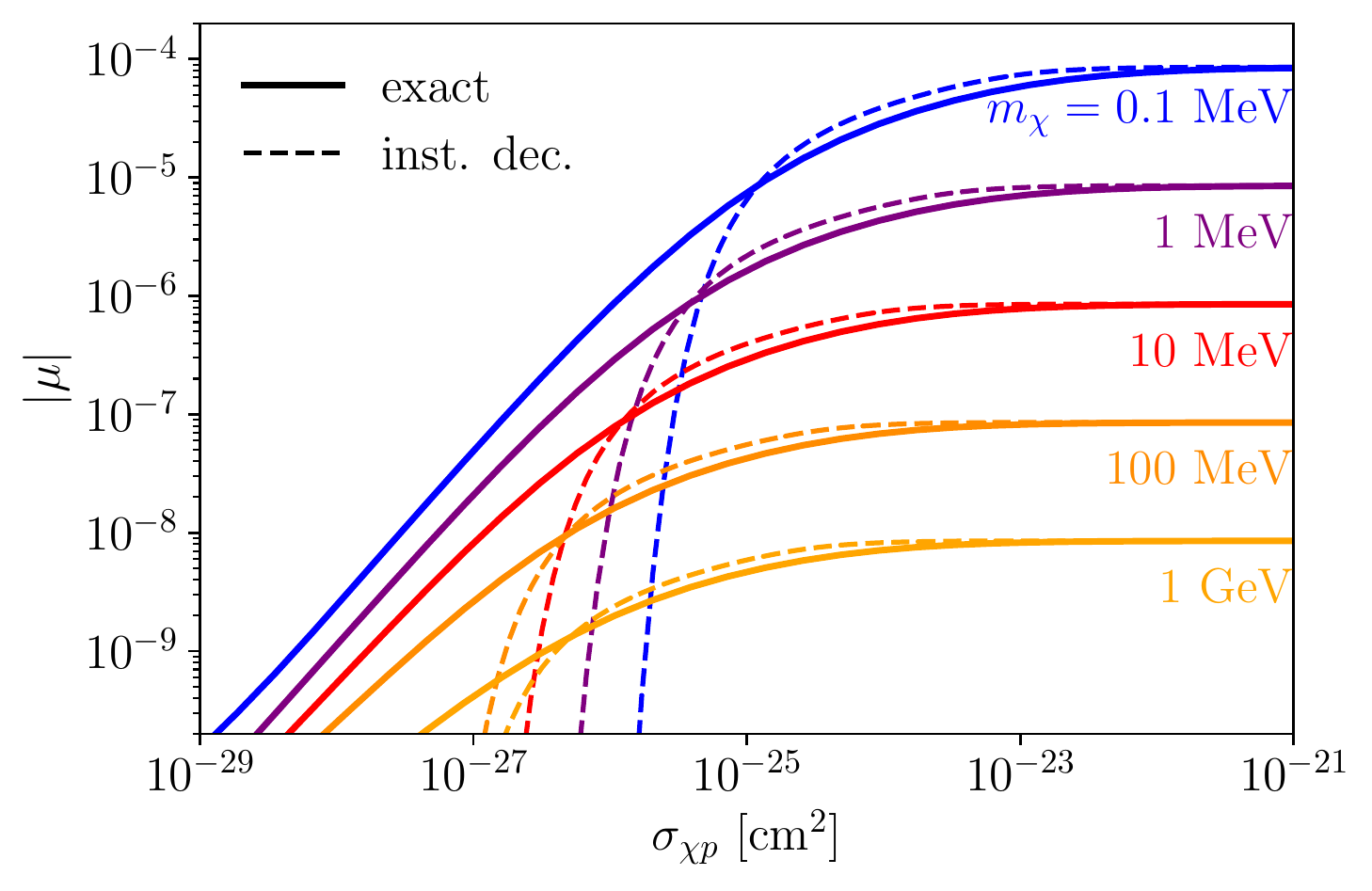}
\caption{Absolute value of the $\mu$-distortion generated by DM-proton scattering with a velocity-independent cross section, for DM masses ranging from 0.1 MeV to 1 GeV. Solid lines show the distortion obtained when explicitly solving the ODE for $T_\chi$ and computing the resulting heat-exchange rate, and dashed lines show the distortion obtained in the instantaneous-decoupling approximation. The latter tends to significantly underestimate $|\mu|$ for low cross sections.}
\label{fig:sigma-mu-ID}
\end{figure}

This is illustrated further in Fig.~\ref{fig:sigma-mu-ID}, where we show the amplitude of the $\mu$-distortion as a function of DM-proton (velocity-independent) cross section. At sufficiently large cross sections, the DM is in thermal equilibrium with protons throughout the $\mu$-era, and the details of its decoupling do not affect the $\mu$ distortion; this explains why the exact and instantaneous-decoupling curves share the same horizontal asymptote at large cross sections, given by Eq.~\eqref{eq:mu-max}. However, in the regime of small cross sections, corresponding to thermal decoupling before the $\mu$-era, the instantaneous-decoupling approximation severely under-estimates the SD amplitude. 

\begin{figure}
\includegraphics[width = 1\columnwidth]{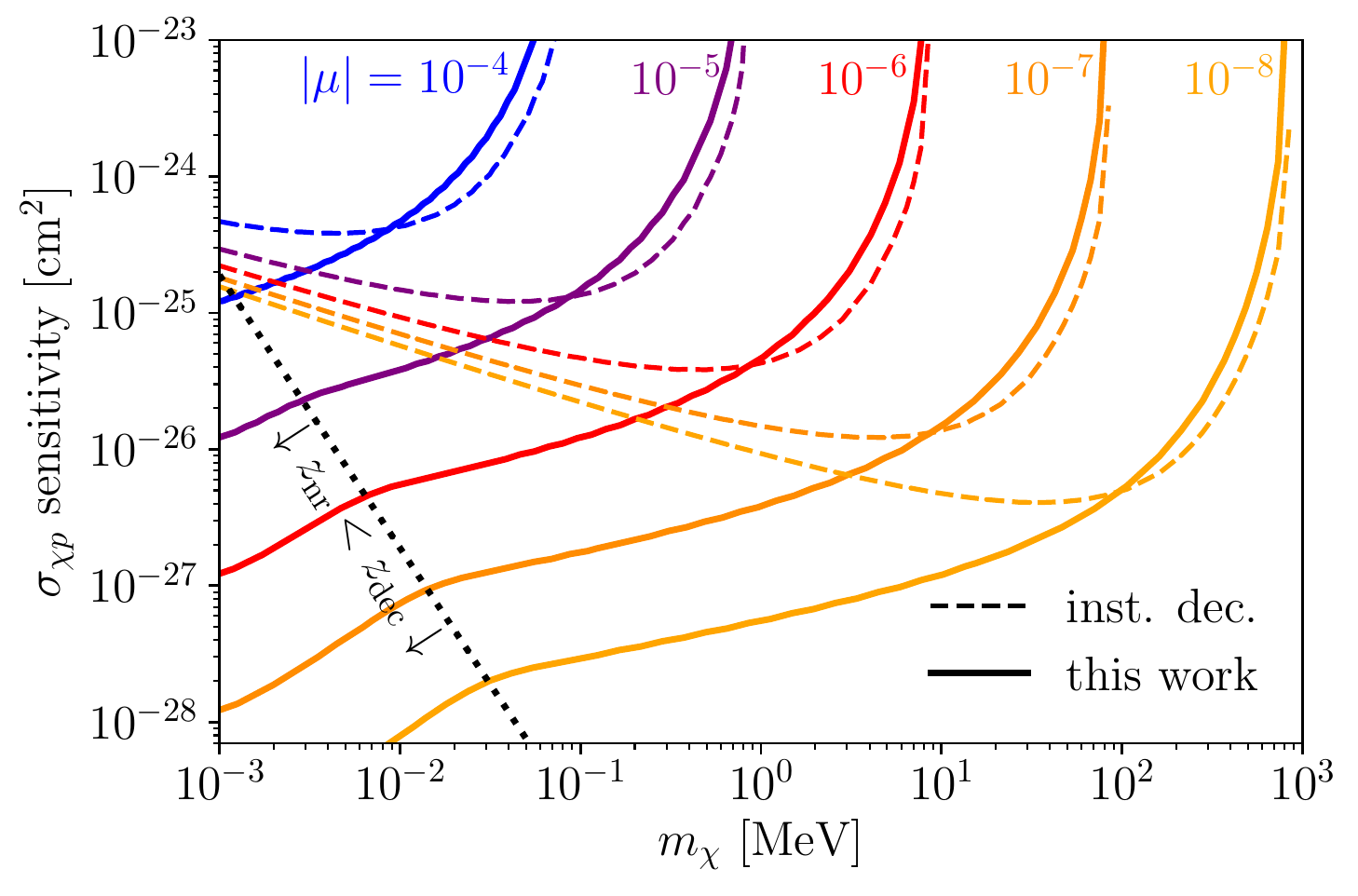}
\caption{Forecasted reach of SD distortion experiments with sensitivity to $|\mu|$ ranging from $10^{-4}$ to $10^{-8}$, for a velocity-independent DM-proton cross section. Solid lines show the sensitivity obtained when solving numerically for the DM temperature evolution and heat-exchange rate, while the dashed lines show the corresponding forecasted reach in the instantaneous-decoupling approximation, used in ACK15. We see that this approximation tends to under-estimate how sensitive SD experiments are to DM interactions. The region below the dotted black line is such that DM thermally decouples before becoming non-relativistic; the change of slope of the forecasted sensitivity to $\sigma_{\chi p}(m_\chi)$ is explained qualitatively at the end of Sec.~\ref{sec:relativistic}.}
\label{fig:mchi-sigma-ID}
\end{figure}

As a consequence, and as illustrated in Fig.~\ref{fig:mchi-sigma-ID}, the estimated sensitivity of a given SD experiment to the DM elastic scattering cross sections tend to be significantly under-estimated in the instantaneous-decoupling approximation, in particular for low DM masses. 

It is interesting to understand the scalings of $\sigma_*(|\mu|; m_\chi)$ in both cases. Using Eq.~(4) of ACK15 for $n = 0$, and in the limit $m_\chi \ll m_p$, we find that the characteristic decoupling scale factor scales as $a_{\rm dec} \propto m_\chi^{1/3} \sigma_*^{2/3}$. In the instantaneous-decoupling approximation, the $\mu$-distortion is exponentially suppressed for $a_{\rm dec} \leq a_\mu$, and approximately scales as $|\mu| \propto \ln(a_{\rm dec}/a_\mu)/m_\chi$ for $a_{\rm dec} > a_\mu$. For very light DM particles, inverting this equation implies $a_{\rm dec} \approx a_\mu$, up to small corrections. Therefore, in the instantaneous-decoupling approximation, we find $\sigma_*(|\mu|; m_\chi) \propto m_\chi^{-1/2}$, almost independent of $|\mu|$; this is indeed the scaling of the dashed lines at low mass in Fig.~\ref{fig:mchi-sigma-ID}. In reality, there is a residual heat-exchange after $a_{\rm dec}$. For $a_{\rm dec} \ll a_\mu$, $T_\chi \ll T_\gamma$ during the $\mu$ era, and therefore $Q \approx \Gamma_{\chi p}/H \propto a^{-1} \sigma_* m_\chi (T_\chi/m_\chi + T_\gamma/m_p)^{1/2}$, where again we assume $m_\chi \ll m_p$. This is a decreasing function of $a$, and the $\mu$-distortion integral is therfore dominated by $a \approx a_\mu$, i.e.~$|\mu| \propto Q(a_\mu)/m_\chi \propto \sigma_* ((a_{\rm dec}/a_\mu)/m_\chi + 1/m_p)^{1/2}$, where we used $T_\chi/T_\gamma \approx a_{\rm dec}/a$ for $a \gtrsim a_{\rm dec}$. In the limit $m_\chi \ll m_p (a_{\rm dec}/a_\mu)$, we thus obtain $|\mu| \propto \sigma_* (a_{\rm dec}/m_\chi)^{1/2} \propto \sigma_*^{4/3} m_\chi^{-1/3}$. This is indeed the behavior seen in the solid lines in Fig.~\ref{fig:sigma-mu-ID}. Inverting this relation, we find $\sigma_*(|\mu|; m_\chi) \propto |\mu|^{3/4} m_\chi^{1/4}$ at low masses, which is indeed the scaling of the solid lines in Fig.~\ref{fig:mchi-sigma-ID}, in the region above the dotted black line.

To conclude this section, the instantaneous-decoupling approximation used in ACK15 turns out to be significantly inaccurate for light DM particles, and the analytic approximations provided in that paper should not be used. Instead, one should solve for the temperature evolution and compute the heat-exchange rate numerically, as we do in the remainder of this work.

\section{Limits and forecasts for DM interactions with a single scatterer}  \label{sec:limits-PL}

Most concrete particle-DM models imply simultaneous interactions with multiple SM particles. Depending on the context, one of these interactions may be dominant. In this section, we consider an idealized DM candidate scattering with one single scatterer at a time.

\subsection{Dark matter-proton scattering}

We start by considering DM-proton interactions, parameterized by a power law $\sigma_{\chi p}(v) \propto v^n$. For positive (and even) exponents $n \geq 0$, such cross sections arise from effective DM-nucleon interactions in the non-relativistic limit \cite{Fitzpatrick_13, Anand_13, Boddy_18}. The case $n = -2$ can arises for a DM particle with an electric dipole moment, in which case DM-electron and DM-photon interactions are necessarily present as well. In this section we restrict ourselves to DM-proton interactions only, but will consider DM with and electric dipole moment self-consistently in Sec.~\ref{sec:dipole}. Moreover, for definiteness, we consider a DM particle that does not interact with Helium nuclei, as would be the case e.g.~for the spin-dependent operators considered in Ref.~\cite{Boddy_18}. Note that our code is general enough to account for DM-He interactions when relevant.

\begin{figure*}
\includegraphics[width = 2\columnwidth]{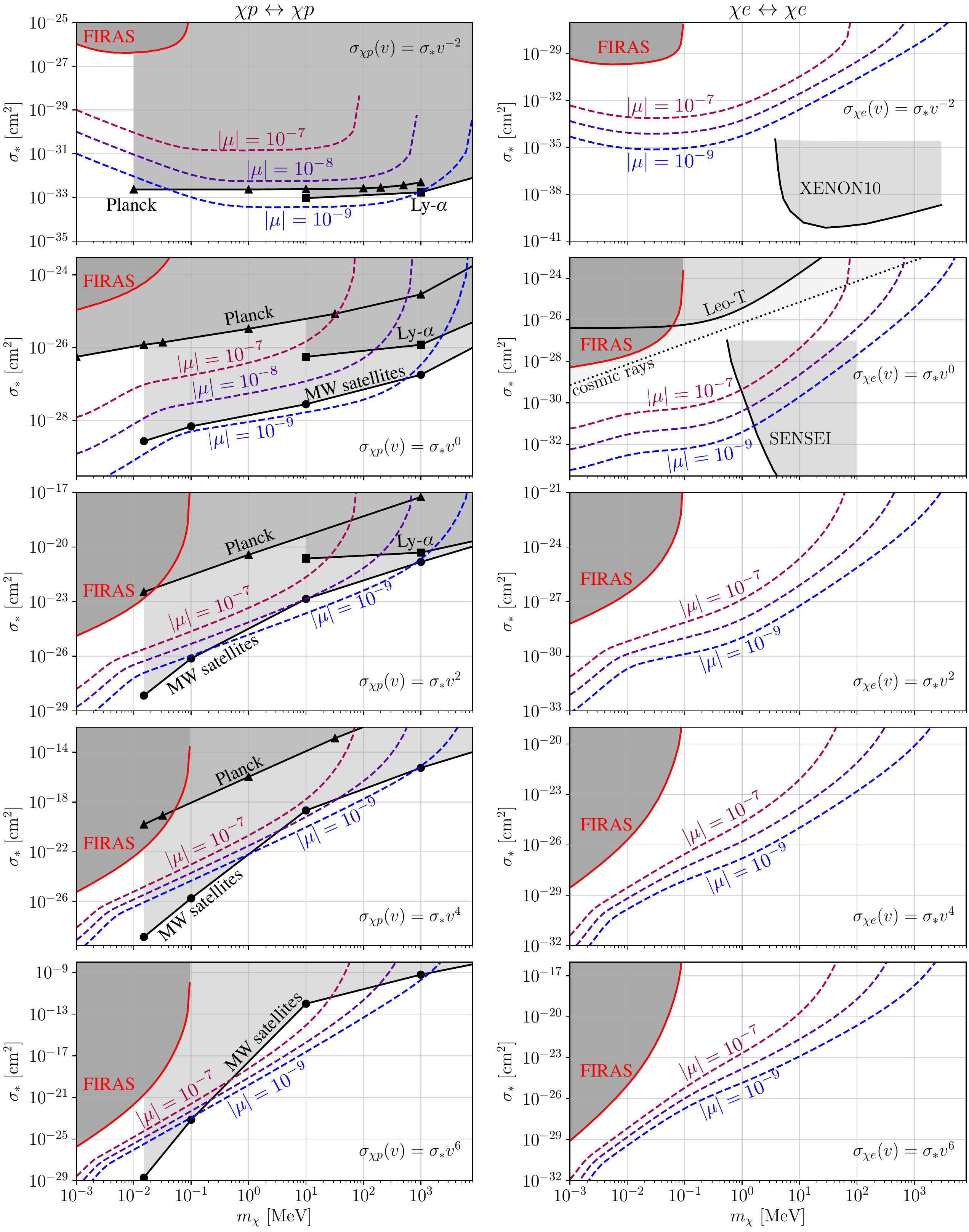}
\caption{FIRAS upper limits (solid red) and forecasted reach of a spectral-distortion experiment with sensitivity to $|\mu|$ as low as $10^{-7}, 10^{-8}$ and $10^{-9}$ (dashed lines), for a DM-proton (left) and DM-electron (right) cross section scaling as $\sigma(v) = \sigma_*v^n$, with $n = -2, 0, 2, 4, 6$ from top to bottom. For DM-proton scattering, when available we show the upper limits from \emph{Planck} temperature and polarization anisotropies \cite{Boddy_18, Boddy_18b} (triangles), Lyman-$\alpha$ forest measurements \cite{Xu_18} (squares), and Milky-Way satellite abundance \cite{Maamari_20} (circles). For DM-electron scattering, we show the cosmic ray reverse direct detection limit of Ref.~\cite{Cappiello_18}, the Leo-T heat-exchange limit of Ref.~\cite{Wadekar_19}, and the direct-detection limits from SENSEI \cite{Abramoff_19} and XENON10 \cite{Essig_12}.}\label{fig:prot-elec}
\end{figure*}

In this paper we focus on light $(m_\chi \lesssim $ GeV) DM particles, for which DM-nuclei interactions are poorly constrained by direct-detection experiments. In this mass range, DM-nuclei interactions are constrained by a variety of astrophysical and cosmological observables: CMB temperature and polarization anisotropies \cite{Dvorkin_14, Boddy_18, Gluscevic_18, Xu_18}, the Lyman-$\alpha$ forest \cite{Xu_18}, the abundance of Milky-Way satellites \cite{Maamari_20}, and the heating/cooling rates in the dwarf galaxy Leo-T and Galactic gas clouds \cite{Wadekar_19}. We show these limits in the left column of Fig.~\ref{fig:prot-elec} (for clarity, we do not show the limits of Ref.~\cite{Wadekar_19} as they are weaker than other existing limits in this mass range). We also show the spectral-distortion limits from FIRAS and the forecasted reach of high-sensitivity SD experiments. We see that an experiment with a sensitivity to $\mu$ of order $\sim 10^{-8} - 10^{-9}$ would be competitive with, or in some cases surpass, the tightest bounds from Milky-Way satellite counts.

\subsection{Dark matter-electron scattering}

It is possible that DM is only coupled to the lepton sector of ordinary matter \cite{Bernabei_08}, in which case DM particles would elastically scatter with electrons, but not nuclei.  

Ref.~\cite{Wadekar_19} provide a constraint from heat exchange in the dwarf galaxy Leo-T. To our knowledge, no linear-cosmology limits were derived for DM-electron interactions. While the fundamental physical processes at play (heat and momentum exchange with the photon-baryon plasma) are identical to those of DM-baryon interactions, the relevant rates depend on the scatterer's mass in a non-trivial fashion -- in particular, the momentum-exchange rates depend on the relative velocity of DM and electrons, whose variance is proportional to $T_\chi/m_\chi + T_b/m_e$. As a consequence, one cannot simply rescale those results to derive DM-electron constraints. 

In addition, DM-electron interactions are constrained by direct-detection experiments for DM masses as low as a few MeV \cite{Essig_12, Essig_17}. For a constant cross section, the strongest limits are from SENSEI \cite{Abramoff_19}. For a cross section scaling as $1/v^2$ (i.e.~form factor $F(q) \propto 1/q$, where $q$ is the momentum transfer), Ref.~\cite{Essig_12} explicitly derives upper limits for DM-electron scattering with the XENON10 experiment. Specifically, their convention is
\beq
\frac{d \sigma_{\chi e}}{d \Omega} = \frac{\overline{\sigma}_e}{4 \pi} \left(\frac{\alpha m_e}{q}\right)^2, 
\eeq
where $q^2 \equiv 2 \mu_{\chi e}^2 v^2 ( 1- \cos \theta)$ and $\mu_{\chi e}$ is the reduced DM-electron mass. Therefore, the momentum-transfer cross section takes the form $\sigma_*/v^2$, where 
\beq
\sigma_* = \frac12 \left(\frac{\alpha m_e}{\mu_{\chi e}}\right)^2 \overline{\sigma}_e.
\eeq

In the right column of Fig.~\ref{fig:prot-elec}, we compare these existing limits to the FIRAS limits on the DM-electron cross section, as well as the forecasted reach of a SD experiment sensitive to $\mu = 10^{-7}, 10^{-8}, 10^{-9}$. Here again, we see that SDs complement existing experiments and can probe regions of parameter space not currently tested. The FIRAS limits we present for $\sigma_{\chi e}(v) \propto v^2, v^4, v^6$, even if limited in their mass reach, appear to be the first constraints derived for these cross sections. 

Note that Ref.~\cite{Cappiello_18} derive limits on DM-electron interactions through their effect on cosmic-ray spectra, assuming a constant DM-electron cross section. Since this limit depends on the cross section at relativistic relative velocities, it cannot be directly compared to other limits, holding in the non-relativistic regime, and we show it as a dashed line in Fig.~\ref{fig:prot-elec}.

\subsection{Dark matter-photon scattering}

All or part of the DM may have an effective coupling to the electromagnetic field, e.g.~through a small (effective) electric charge \cite{Dubovsky_01, Dubovsky_04}, an electric or magnetic dipole moment \cite{Sigurdson_04},  or other effective coupling operators \cite{Weiner_12}. Clearly, any coupling to the standard photon implies indirect couplings to nuclei and electrons; in this section, for simplicity, we consider a DM interacting only with photons, with a power-law cross section of the form \eqref{eq:sigma-photon}.

\begin{figure}
\includegraphics[width = \columnwidth]{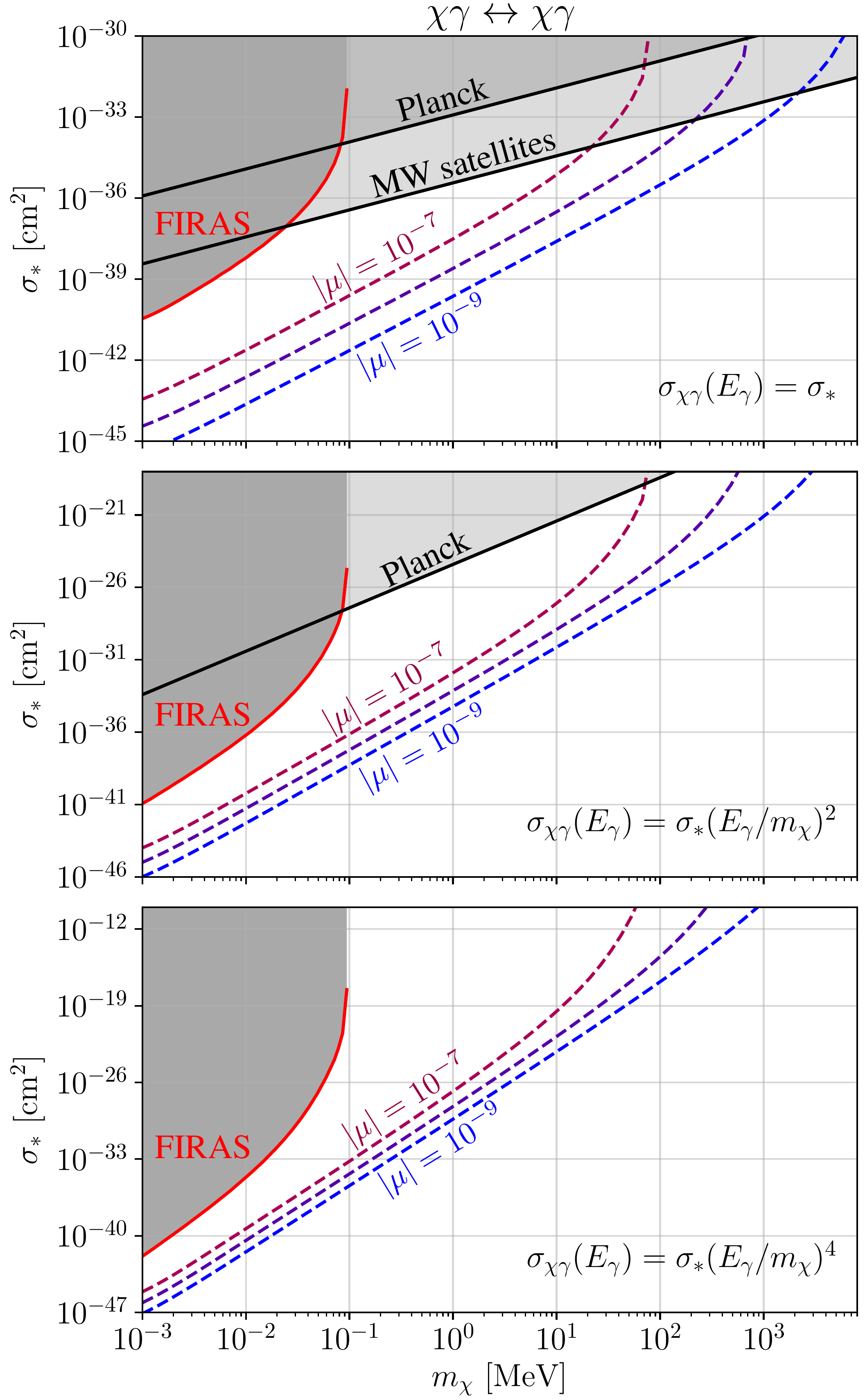}
\caption{Upper limits on a DM-photon cross section scaling as $E_\gamma^0$ (upper panel), $E_\gamma^2$ (middle panel) and $E_\gamma^4$ (lower panel). The Planck limit for a constant cross section is from Ref.~\cite{Becker_20}, and that for $\sigma_{\chi \gamma} \propto E_\gamma^2$ is obtained from Ref.~\cite{Wilkinson_14a}. The limit derived from the abundance of Milky-Way satellites is from Ref.~\cite{Schewtschenko_16}. In addition, we show the spectral-distortion limits from FIRAS, resulting from the upper bound $|\mu| < 9 \times 10^{-5}$, as well as the forecasted reach of a spectral-distortion experiment with sensitivity to $|\mu|$ of $10^{-7}, 10^{-8}, 10^{-9}$.} \label{fig:photon_lims}  
\end{figure}

Upper limits on a constant DM-photon cross section were derived in Refs.~\cite{Wilkinson_14a, Stadler_18, Becker_20} using CMB-anisotropy data and in Refs.~\cite{Boehm_14, Schewtschenko_16} from the abundance of Milky-Way satellites. We compare them with the FIRAS upper limit and the forecasted sensitivity of future SD experiments in the top panel of Fig.~\ref{fig:photon_lims}. We see that FIRAS sets more stringent constrains at very low masses $m_\chi \lesssim 20$ keV, and that more sensitive SD experiments would be able to probe DM-photon elastic cross sections much weaker than the current upper limits, over a broad range of masses. 

Ref.~\cite{Wilkinson_14a} also derived an upper limit on a DM-photon (weighted) cross section scaling as temperature squared. Simple kinematic considerations show that the weighted integral of the cross section appearing in the momentum-exchange rate relevant to CMB-anisotropy limits is precisely $\langle \sigma_{\chi \gamma} \rangle(T_\gamma)$ given in Eq.~\eqref{eq:sigma_gamma-av}. Thus $\langle \sigma_{\chi \gamma} \rangle(T_\gamma) \propto T_\gamma^2$ fundamentally corresponds to a momentum-exchange cross section $\sigma_{\chi \gamma}(E_\gamma) \propto E_\gamma^2$. Assuming the specific form $\sigma_{\chi \gamma}(E_\gamma) = \sigma_* (E_\gamma/m_\chi)^2$, we then get $\langle \sigma_{\chi \gamma} \rangle(T_\gamma) = \frac{20 \pi^2}{7} \sigma_* (T_\gamma/m_\chi)^2$. Assuming the interacting particle makes all of the DM, Ref.~\cite{Wilkinson_14a} derive the following constraint on the weighted cross section evaluated at today's temperature: $\langle \sigma_{\chi \gamma} \rangle(T_0) \leq 6 \times 10^{-40} (m_\chi/$GeV) cm$^2$. We compare this limit against the FIRAS limit and the forecasted sensitivity of future SD experiments in the middle pannel of Fig.~\ref{fig:photon_lims}. We see that the FIRAS limits are significantly stronger than Planck limits for $m_\chi \lesssim 0.1$ MeV; this is because of the relatively stronger interactions at high redshift, due the quadratic energy scaling. We also find that a SD experiment with a sensitivity $|\mu| \sim 10^{-8}$ would be able to detect a cross section orders of magnitude weaker than Planck's current limits, for DM masses as large as $m_\chi \sim$ GeV.

Lastly, for completeness, in the bottom panel of Fig.~\ref{fig:photon_lims}, we show FIRAS limits and the forecasted sensitivity of future SD experiments for a DM-photon cross section scaling as $E_\gamma^4$, for which no other constraints exist to our knowledge (although constraints on a specific model leading to such an interaction are discussed in Ref.~\cite{Weiner_12}).

\section{Application to a DM particle with a dipole moment} \label{sec:dipole}

We now consider a specific particle-DM model, in which DM can simultaneously interact with photons, electrons and nuclei: a DM particle with an electric or magnetic moment. The cosmological implications of such a DM model were first studied in Ref.~\cite{Sigurdson_04}. After reviewing the different interactions and their cross sections in Sec.~\ref{sec:dipole-cross-sec}, we summarize existing constraints on the DM dipole moments (translating model-independent constraints when applicable) in Sec.~\ref{sec:dipole-constraints}. We then compute the upper bounds on the DM dipole moments resulting from FIRAS constraints, and forecast the reach of future SD experiments in Sec.~\ref{sec:dipole-SD}. Throughout this section we work in natural units, $\hbar = c = 1$ and denote by $\alpha \equiv e^2/4 \pi$ the fine-structure constant, where $e$ is the elementary charge.

\subsection{Cross sections} \label{sec:dipole-cross-sec}

We consider a DM particle with either an electric dipole moment $\mathcal{D}$ or a magnetic dipole moment $\mathcal{M}$ (but not both simultaneously). The dipole moments have dimensions of inverse mass, and we will work with the following dimensionless parameters 
\beq
\alpha_E \equiv \frac{\mathcal{D} m_\chi}{e}, \ \ \ \ \ \ \ \ \alpha_M \equiv \frac{\mathcal{M} m_\chi}{e}.
\eeq 
Refs.~\cite{Sigurdson_04, Fortin_12} argue on simple dimensional grounds that $\alpha_E$ and $\alpha_M$ shoud be less than unity. In more detail, such dipole moments would arise from loops involving heavy charged particles coupled to $\chi$, giving $\alpha_E, \alpha_M \sim g^2( m_\chi/M) \lesssim 1$, where $g \lesssim 1$ is the coupling between $\chi$ and the heavy charged particle of mass $M \gtrsim m_\chi$ \cite{Graham_12, Chang_19}.

For either an electric or a magnetic dipole moment, the momentum-exchange DM-photon cross section is \cite{Gell-Mann_54, Sigurdson_04}
\barr
\sigma_{\chi \gamma}(E_\gamma) =  \frac{64\pi}{3 m_\chi^2} \alpha^2  \alpha_\chi^4 \left(\frac{E_\gamma}{m_\chi}\right)^2 \ \ \ [\alpha_\chi = \alpha_E \ \textrm{or} \  \alpha_M].~~~\label{eq:sigma-gamma-dipole}
\earr
The momentum-exchange cross sections for elastic scattering with scatterers  $s$ with charge $Z_s e$ can be derived from the differential cross sections provided in Refs.~\cite{Sigurdson_04, Sigurdson_04_erratum}: 
\barr
\sigma_{\chi s}(v) &=& \frac{8 \pi Z_s^2}{m_\chi^2}  \frac{\alpha^2 \alpha_E^2 }{v^2}, \\
\sigma_{\chi s}(v)  &=& \frac{8 \pi Z_s^2}{m_\chi^2}R_{\chi s}~ \alpha^2 \alpha_M^2, 
\earr
for an electric or magnetic dipole, respectively, where in the latter case we used
\beq
R_{\chi s} \equiv \frac{m_s^2 + m_s m_\chi + \frac32 m_\chi^2}{(m_s + m_\chi)^2} \in (5/6, 3/2).
\eeq
In addition, a particle with an electric or magnetic dipole moment annihilates to photons, with cross section \cite{DelNobile_12, Fortin_12}
\beq
\sigma_{\chi \chi\rightarrow \gamma\gamma} ~v  = \frac{4 \pi}{m_\chi^2} \alpha^2 \alpha_\chi^4 \ \ \ \ \   [\alpha_\chi = \alpha_E \ \textrm{or} \  \alpha_M]. \label{eq:EDM_ann_gg}
\eeq
It also annihilates to unit-charged fermion-antifermion pairs with cross sections \cite{Sigurdson_04_erratum, Masso_09, Fortin_12, DelNobile_12}
\barr
\sigma_{\chi \chi\rightarrow f \overline{f}} ~ v  &\approx& N_{c, f} ~\frac{\pi}{3 m_\chi^2} \alpha^2 \alpha_E^2 ~v^2, \label{eq:sigma_ann_f_E} \\
\sigma_{\chi \chi\rightarrow f \overline{f}} ~ v  &\approx& N_{c, f} \frac{4 \pi}{m_\chi^2} \alpha^2 \alpha_M^2,\label{eq:sigma_ann_f_M}
\earr
for an electric or magnetic dipole, respectively, where $N_{c, f}$ is the number of color degrees of freedom for fermions ($N_{c, f} = 1$ for leptons and 3 for quarks). Note that Eqs.~\eqref{eq:sigma_ann_f_E}-\eqref{eq:sigma_ann_f_M} hold in the limit $m_\chi \gg m_f$. 

\subsection{Prior constraints} \label{sec:dipole-constraints}

\begin{figure*}
\includegraphics[width = \columnwidth]{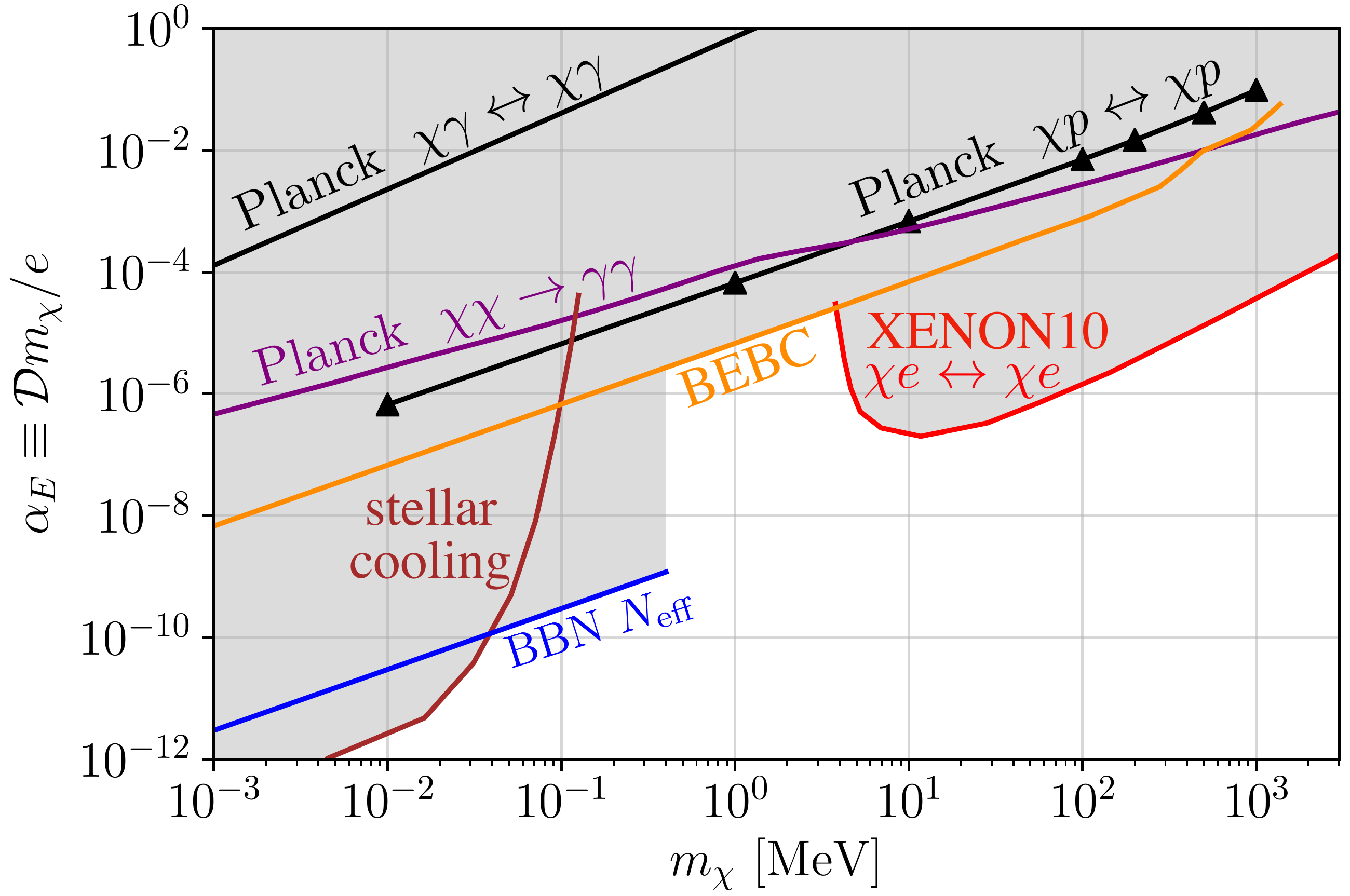}
\includegraphics[width = \columnwidth]{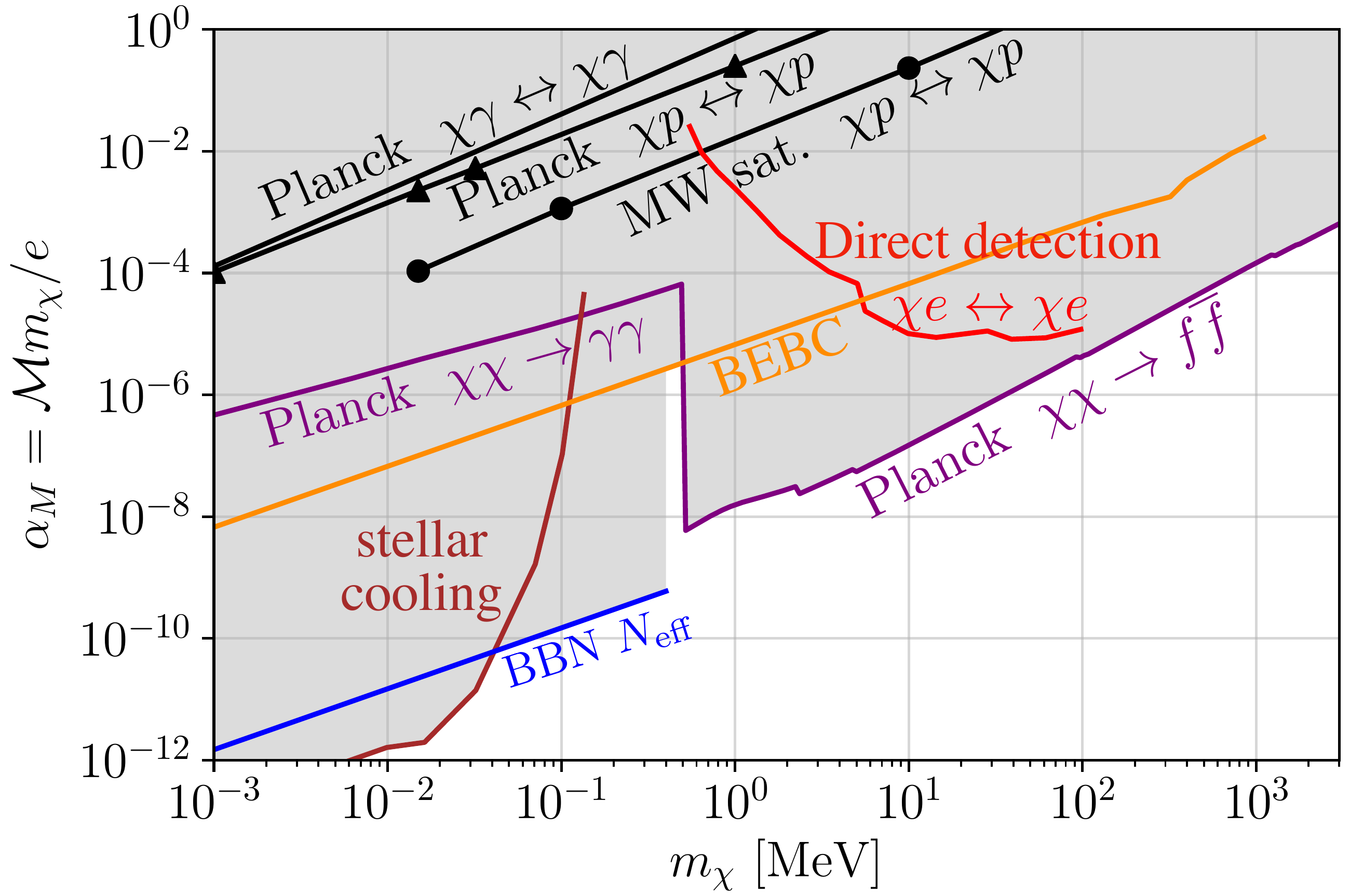}
\caption{Constraints on the DM dimensionless electric (left) and magnetic (right) dipole moments, either taken directly from previous works, or derived from model-independent constraints. The Planck elastic-scattering bounds are derived from the limits of Ref.~\cite{Wilkinson_14a} on DM-photon scattering and from the limits of Refs.~\cite{Boddy_18, Boddy_18b} for DM-baryon scattering. The Milky-Way satellite limits are derived from Ref.~\cite{Maamari_20}. The Planck annihilation limits are derived from Ref.~\cite{Planck_18}. The stellar cooling and BBN limits are taken from Ref.~\cite{Chang_19} (see also \cite{Chu_19b}). The BEBC limit is from Ref.~\cite{Marocco_20}, and the direct-detection limits are derived from Refs.~\cite{Essig_12} for the electric dipole moment and from \cite{Abramoff_19} for the magnetic dipole moment.}\label{fig:dipole_existing}
\end{figure*}

We now summarize and comment on non-SD constraints on the DM dipole moments and show them in Fig.~\ref{fig:dipole_existing}. Whenever relevant, we assume the particle $\chi$ makes all of the DM. See also Fig.~10 of Ref.~\cite{Chu_19a} and Fig.~5 of Ref.~\cite{Chu_19b} for similar compilation of constraints, with different conventions.

$\bullet$ First, we translate the single-scatterer constraints discussed in Sec.~\ref{sec:limits-PL} into upper bounds on $\alpha_\chi = \alpha_E, \alpha_M$. For both electric and magnetic dipole moments, we obtain a bound from the CMB-anisotropy limit on DM-photon scattering with cross section quadratic in temperature \cite{Wilkinson_14a}. For the electric dipole case, we may translate Refs.~\cite{Boddy_18b}'s upper limit on $\sigma_{\chi p} \propto 1/v^2$. For the magnetic dipole case, we conservatively use the spin-dependent limits on a constant DM-proton cross section, i.e.~we neglect interactions with helium nuclei, as they do not have the same cross section for a magnetic-dipole DM and for the effective spin-independent operator considered in Ref.~\cite{Boddy_18}. We see that the bound on $\alpha_E$ inferred from CMB-anisotropy limits is stronger than that on $\alpha_M$, which is due to the $1/v^2$ enhancement of the DM-proton cross section for an electric dipole moment. We also show the bounds derived from Milky-Way satellite limits on DM-proton scattering \cite{Maamari_20}, as well as those derived from direct-detection limits on DM-electron scattering \cite{Essig_12, Abramoff_19}. For clarity, we do not show the upper limit inferred from the Lyman-$\alpha$ limits on DM-proton scattering \cite{Xu_18}; they are comparable to the Planck limits for $\alpha_E$, and lie between the Planck and Milky-Way satellite limits for $\alpha_M$. 

$\bullet$ CMB anisotropies are a very sensitive probe of energy injection, through its effect on cosmological recombination \cite{Adams_98, Chen_04}. In particular, the latest Planck observations \cite{Planck_18} constrain the DM annihilation cross section to 
\beq
f_{\rm eff} \frac{\langle \sigma_{\rm ann} v \rangle}{m_\chi} \leq 3.2  \times 10^{-28} ~ \textrm{cm}^3 ~\textrm{s}^{-1} \textrm{GeV}^{-1}, \label{eq:Planck-ann}
\eeq
where $f_{\rm eff}$ is the effective deposition efficiency at $z \lesssim 10^3$ relevant to CMB anisotropies. This constraint can be translated into bounds on $\alpha_\chi = \alpha_E, \alpha_M$ \cite{DelNobile_12}. Annihilations to photons and electron-positron pairs have the highest effective deposition efficiency, which we obtain from Ref.~\cite{Slatyer_16}. For other fermions-antifermions, we conservatively assume a constant $f_{\rm eff} = 0.15$.

Note that the constraint \eqref{eq:Planck-ann} specifically applies to $s$-wave annihilations, i.e.~$\sigma_{\rm ann}v $ = constant. For $p$-wave annihilations, i.e.~$\sigma_{\rm ann} v \propto v^2$, Ref.~\cite{Liu_16} also derive constraints from the heating of the intergalactic medium, by estimating the DM phase-space properties in the first non-linear structures. We find these limits to be much less constraining for the electric dipole moment than the $s$-wave limits, and do not include them in Fig.~\ref{fig:dipole_existing}.

$\bullet$ If the DM is thermally produced before Big Bang nucleosynthesis (BBN), masses $m_\chi \lesssim 1$ MeV are strongly disfavored by the upper limit on the effective number of neutrinos \cite{Mangano_11}. Ref.~\cite{Chang_19} computes the maximum coupling above which DM reaches chemical equilibrium with the SM bath. With the most conservative assumption of a reheating temperature of $10$ MeV, they find $\mathcal{D} \lesssim 9 \times 10^{-4}$ TeV$^{-1}$ and $\mathcal{M} \lesssim 5 \times 10^{-4}$ TeV$^{-1}$. We show these limits for $m_\chi \leq 0.4$ MeV, which is the most conservative constraint of  Ref.~\cite{Sabti_20} on the mass of thermally-produced particles. 

$\bullet$ In addition to these cosmological bounds, collider constraints were derived in Ref.~\cite{Fortin_12}. Fixed-target proton experiments also constrain the DM dipole moments \cite{Chu_20}; in particular, we show the recent Big European Bubble Chamber (BEBC) constraint of Ref.~\cite{Marocco_20}, which is stronger than most other collider constraints in the relevant mass range. Lastly, stellar cooling considerations constrain the electric and magnetic dipole moment \cite{Chu_19b, Chang_19} for $m_\chi \lesssim 0.1$ MeV. 

\subsection{FIRAS constraints and reach of future SD experiments} \label{sec:dipole-SD}

We compute the $\mu$-distortion resulting from simultaneously scattering with photons, electrons and nuclei, as described in Secs.~\ref{sec:physics}-\ref{sec:numerics}. Note that this process systematically leads to a \emph{negative} chemical potential, as the DM extracts heat from the photon-baryon plasma.

In addition, DM annihilations into photons and fermion-antifermion pairs result in a \emph{positive} $\mu$-distortion \cite{McDonald_00, Chluba_13b}, which we compute as follows. Since the redshifts of injection relevant to $\mu$-distortion are $z \gtrsim 5 \times 10^4$, we may safely assume that the energy is entirely deposited on the spot, with 100\% efficiency. The volumetric rate of energy injection is then
\barr
\dot{\mathcal{Q}} &=& \frac12 n_\chi^2\langle \sigma_{\rm ann} v \rangle 2 m_\chi = f_\chi^2 \rho_c^2 \frac{\langle \sigma_{\rm ann} v \rangle}{m_\chi},\\
\sigma_{\rm ann} &\equiv& \sigma_{\chi \chi \rightarrow \gamma \gamma} + \sum_{m_f < m_\chi}  \sigma_{\chi \chi \rightarrow f \overline{f}}
\earr
During the $\mu$-era, the Universe is radiation-dominated hence $H(a) \propto a^{-2}$. As a consequence, the ratio
\beq
\frac{\rho_c^2}{H\rho_\gamma} \approx 2.9 \times 10^{17} ~\textrm{GeV~s/cm}^3 
\eeq
is nearly constant throughout the relevant epoch. The $\mu$-parameter is therefore
\barr
\mu &=& f_\chi^2  \int \frac{da}{a} \mathcal{G}_{\mu}(a)  \frac{\rho_c^2}{H \rho_\gamma} \langle \sigma_{\rm ann} v \rangle/m_\chi \nonumber\\
&\approx& 4\times 10^{-10} \int \frac{da}{a} \frac{\mathcal{G}_{\mu}(a)}5   \frac{ f_\chi^2 \langle  \sigma_{\rm ann} v \rangle/m_\chi}{3 \times 10^{-28} \textrm{cm}^3/\textrm{s/GeV}}.
\earr
In the second line we have normalized $\mathcal{G}_\mu(a)$ to its integral over $d\ln a$ and the annihilation cross section to the Planck CMB anisotropy 95\% upper limit \cite{Planck_18}, holding for a constant $\langle \sigma_{\rm ann} v \rangle$  and for maximum deposition efficiency. We therefore see that for a constant $\langle \sigma_{\rm ann} v \rangle$, only the most sensitive SD experiments can be competitive with CMB anisotropies in terms of their sensitivity to annihilating DM particles. For $p$-wave annihilation, however, spectral distortions could be more constraining than CMB anisotropies.

For a DM particle with either an electric or magnetic dipole moment, we compute the net $\mu$ distortion resulting from elastic scattering with plasma particles, as well as annihilations. We self-consistently use the DM temperature evolved as described in Secs.~\ref{sec:physics}-\ref{sec:numerics} to compute the $p$-wave annihilation cross section, with $\langle v^2 \rangle = 6 T_\chi/m_\chi$.

\begin{figure*}
\includegraphics[width = \columnwidth]{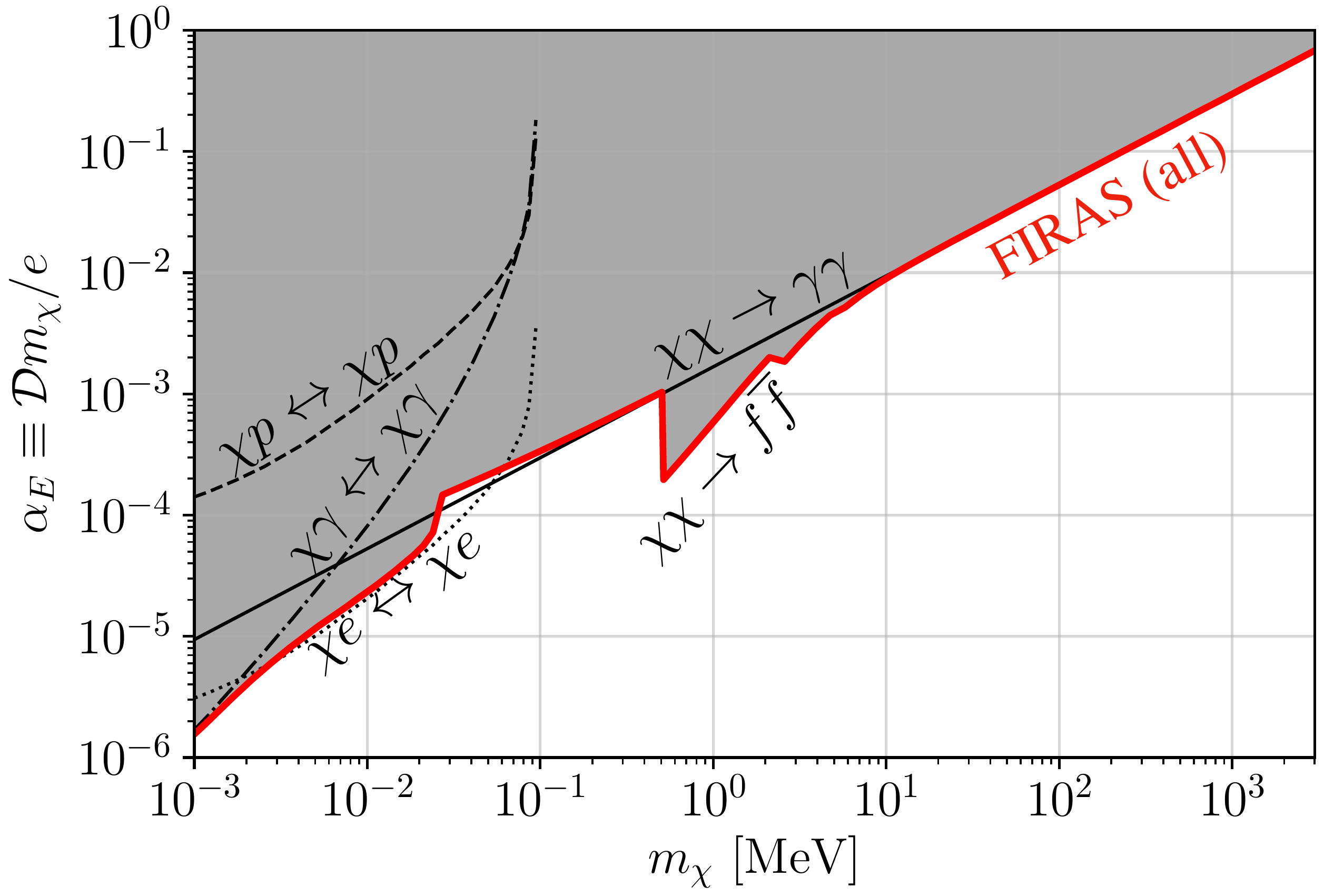}
\includegraphics[width = \columnwidth]{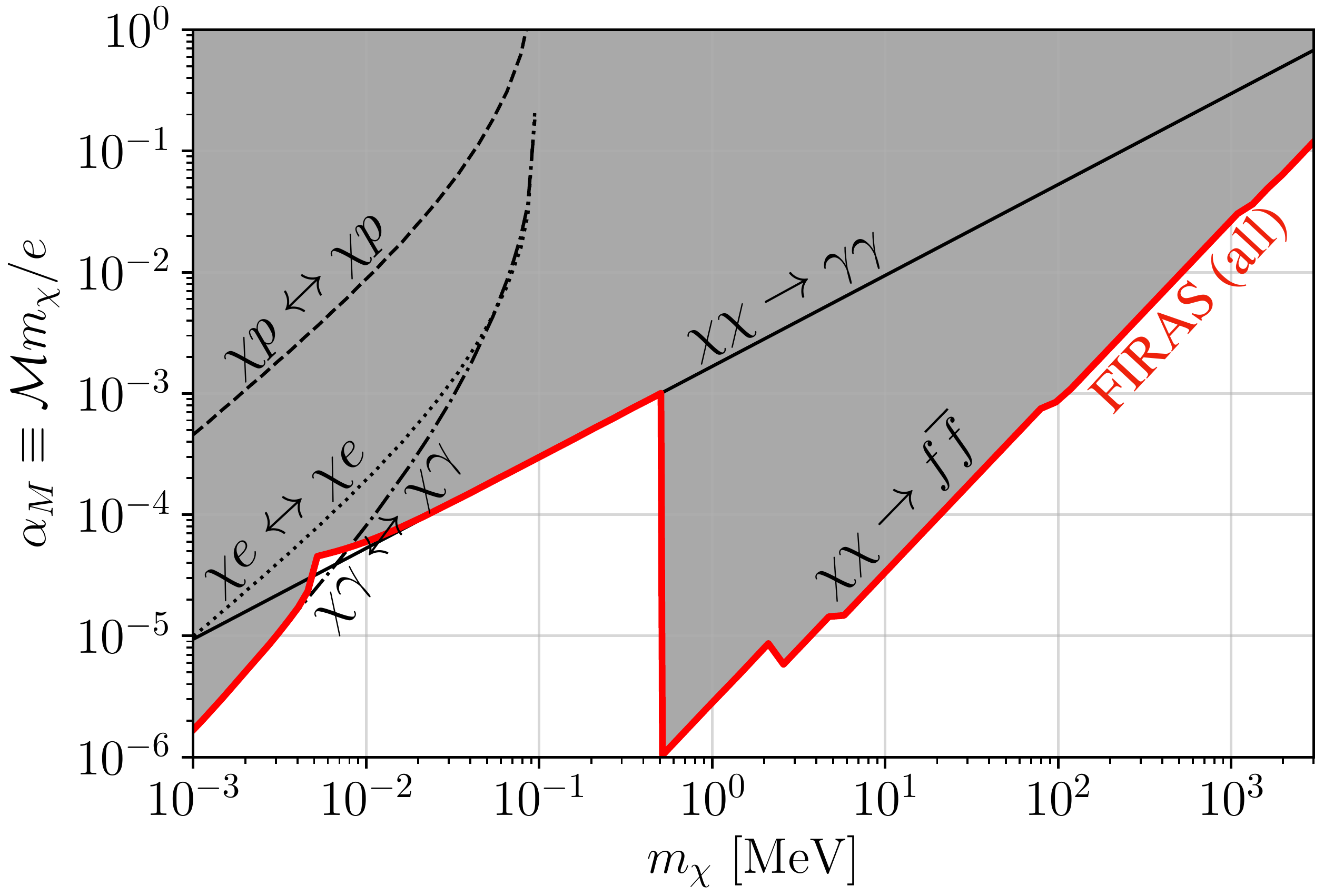}
\caption{FIRAS upper limits on the dimensionless electric (left) and magnetic (right) DM dipole moment. The thin black lines show the limits corresponding to individual processes accounted for separately, and the thick red line shows the overall limit, accounting for all processes simultaneously. The discontinuities at $m_\chi > m_e$ are due to our approximate step-function expression for the cross section for annihilations into fermion-antifermions pairs.}\label{fig:dipole_details}
\end{figure*}

We start by showing the FIRAS upper limits on the dimensionless dipole moments in Fig.~\ref{fig:dipole_details}, as well as the limits that would result from each one of the energy injection/extraction processes accounted for separately. For very low masses $m_\chi \lesssim$ few keV, the upper limits are mostly determined by elastic scattering with photons. In the case of the electric dipole moment, for 2 keV $\lesssim m_\chi \lesssim 20$ keV, the limit is mostly determined by elastic scattering with electrons. For masses greater than a few tens of keV, the upper limits are driven by annihilations into photon pairs. For $m_\chi \geq m_e$, the limits are tightened due to annihilations to electron-positron pairs. This feature is much more pronounced for a magnetic moment, due to the $s$-wave character of annihilations to fermion-antifermion pairs, in contrast to $p$-wave annihilations for an electric dipole moment. The discontinuity of the limit at $m_\chi = m_e$ and at fermion masses is due to our approximate step-function cross section. Note that there is a region where the overall upper limit is \emph{worse} than the limit obtained when considering elastic scattering alone, or annihilations to photons alone. This is because the two processes give rise to $\mu$ distortions with an opposite sign (negative for the former, positive for the latter), and therefore nearly cancel out when they are comparable in magnitude. The qualitative features discussed in the context of FIRAS limits also hold for more sensitive SD experiments: at low masses, SD bounds are determined by elastic scattering, while at high masses, DM annihilations contribute most of the energy injection, thus dominate the limits.

\begin{figure*}
\includegraphics[width = \columnwidth]{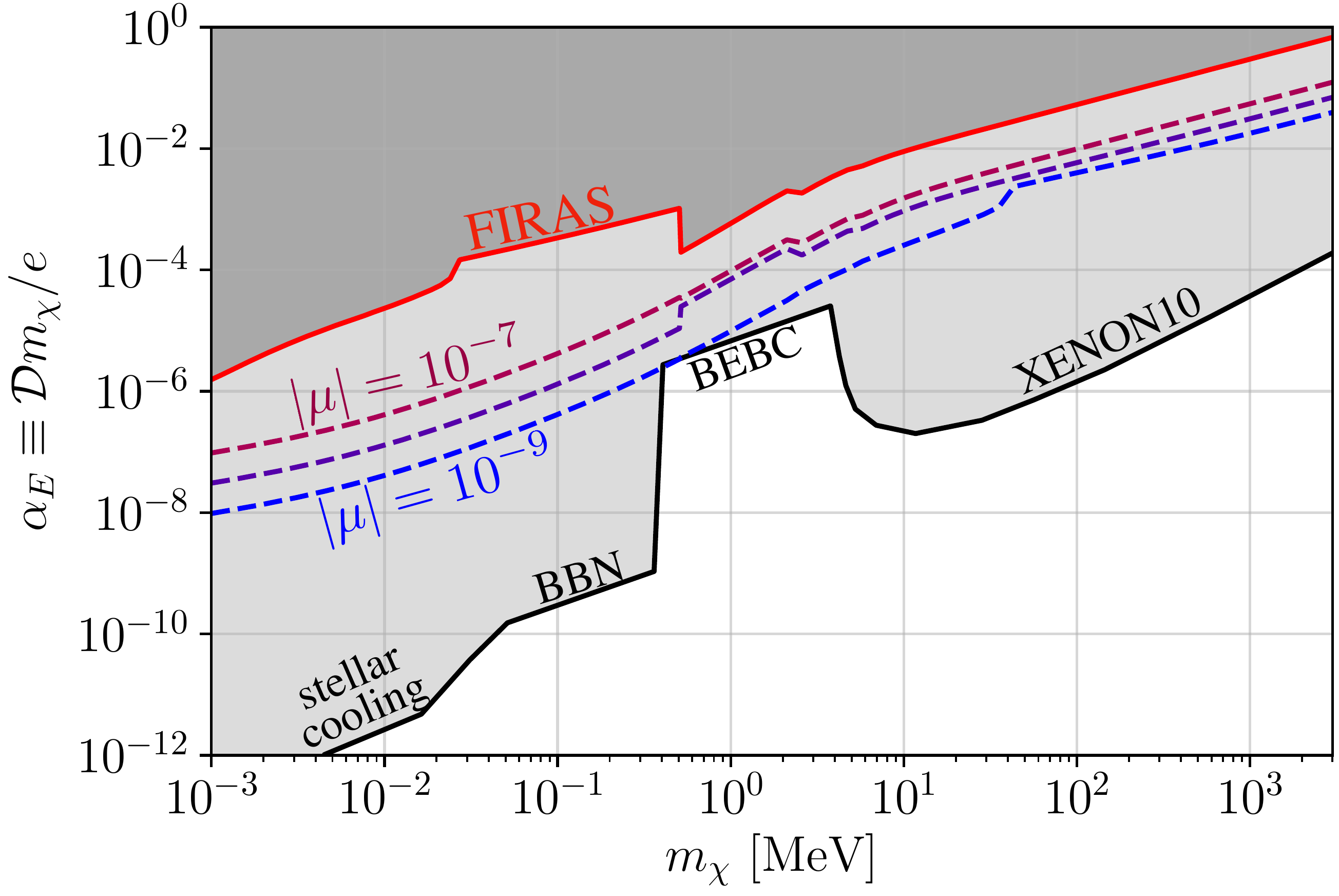}
\includegraphics[width = \columnwidth]{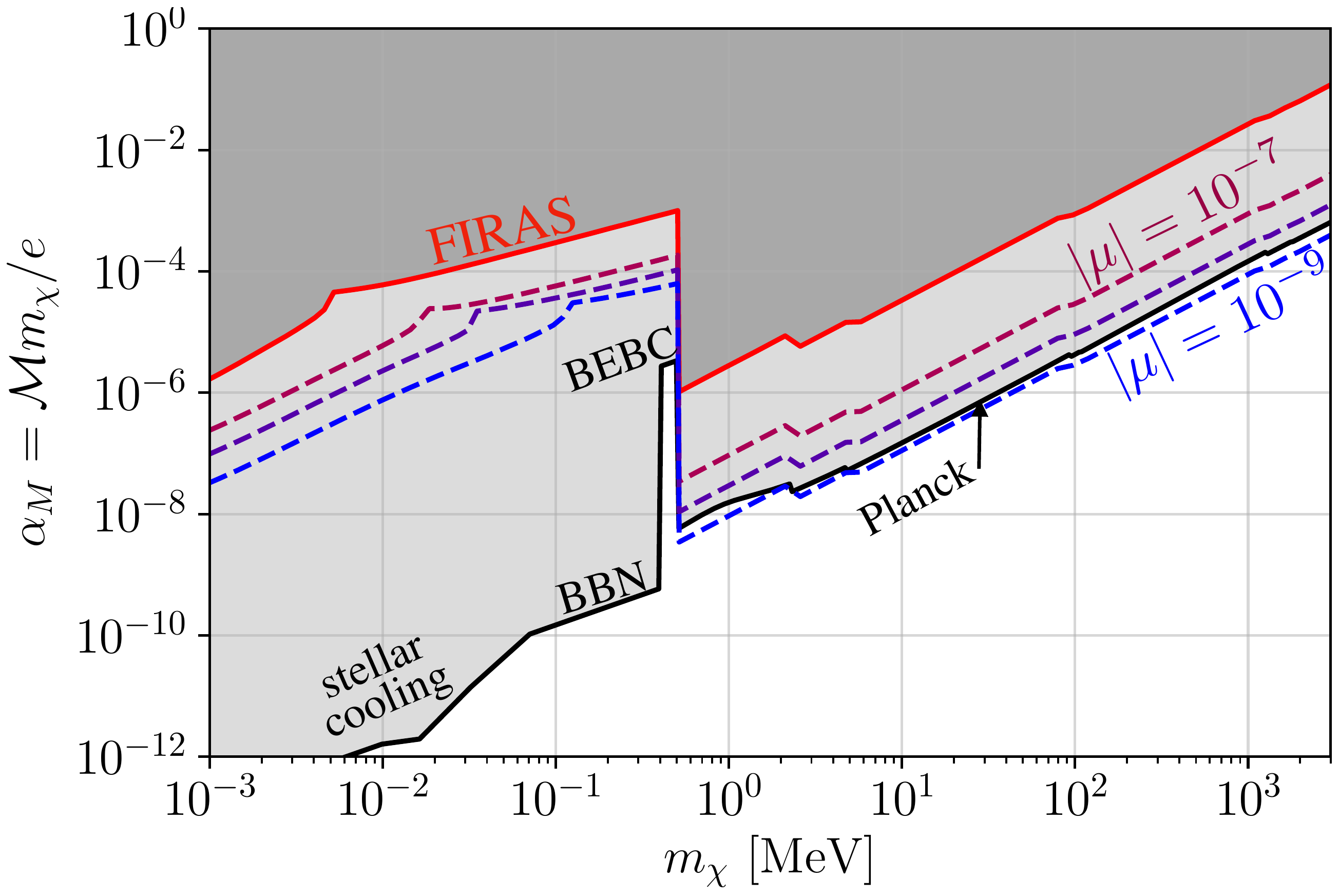}
\caption{FIRAS limits (solid red) and forecasted reach of future SD experiments on the DM dimensionless electric (left) and magnetic (right) dipole moments, for a sensitivity to $|\mu|$ ranging from $10^{-7}$ to $10^{-9}$ (colored dashed lines). The shaded area delimited by the solid black line shows the envelope of other constraints, shown in more detail in Fig.~\ref{fig:dipole_existing}. SD experiments will not be able to reach electric dipole moments below the combined current limits. For $m_\chi \geq m_e$, futuristic SD experiments with sensitivity $|\mu| \approx 10^{-9}$ would in principle be able to probe a magnetic dipole moment just below the current Planck upper limits resulting from DM annihilation constraints.}\label{fig:dipole_SD}
\end{figure*}

Finally, in Fig.~\ref{fig:dipole_SD}, we show the forecasted sensitivity of future SD experiments, as well as the FIRAS limits, overlaid with the strongest constraints for any given mass range. Fig.~\ref{fig:dipole_SD} shows that, even though SD experiments can set interesting limits on their own, they are in general not as sensitive as the best existing limits for any given DM mass, even for an experiment able to probe $|\mu| \approx 10^{-9}$. Only for the magnetic-dipole DM could futuristic SD experiments be competitive with Planck for $m_\chi \geq m_e$, where the tightest current bounds on $\alpha_M$ result from CMB constraints on the DM annihilation cross section.

\section{Conclusions} \label{sec:conclusion}

In this paper, we have revisited our first calculation \cite{ACK15} (ACK15)
of CMB spectral distortion due to DM scattering with standard particles in the early Universe. Importantly, we showed that the detailed numerical evolution of the DM temperature and of the resulting heat-exchange rate leads to significantly larger spectral distortions than the simple instantaneous-decoupling approximation of ACK15. Indeed, light DM particles can have a significant residual heat exchange with the plasma long after their formal thermal decoupling. The analytic approximations of ACK15 are therefore overly conservative and should not be used. Instead, we provide a public code, \codename, which accurately compute the DM thermal evolution and resulting spectral distortion. \codename~takes as inputs the different cross sections of a DM particle (possibly making a fraction of the total DM abundance) with photons, electrons and nuclei, as well as its annihilation cross sections to photons and fermion-antifermion pairs, and computes the $\mu$-distortion jointly produced by all these interactions.

As an illustration, we computed the FIRAS upper limits on the DM cross sections with single scatterers (either photons, electrons or protons), for different energy and velocity dependences. For scattering with electrons in particular, the FIRAS limits presented here are often the only available limits  in the sub-MeV mass range. For scattering with photons, we find that FIRAS is more constraining than CMB-anisotropy and Milky-Way satellite limits at masses $m_\chi \lesssim 30-100$ keV, depending on the energy dependence of the cross section. We also forecasted the reach of future SD experiments to these interactions, and found that they can be competitive with, or in some cases surpass, the best existing upper limits on DM elastic scattering with individual scatterers. 

Realistic DM candidates are expected to have interactions with multiple Standard Model particles. As an example, we computed the SD generated by a DM particle with an electric or magnetic dipole moment, accounting for both elastic scattering and annihilation processes. We computed the resulting FIRAS constraints and forecasted reach of future SD experiments. While SDs on their own do set non-trivial limits on the DM dipole moments, we find that even futuristic SD experiments would not improve on the best available limits on the electric dipole moment of DM. The best current bounds on the magnetic dipole moment of DM particles heavier than the electron result from the Planck annihilation cross section limits; we show that very sensitive SD experiments (reaching $\mu \sim 10^{-9}$) could in principle be competitive with this limit. However, our realistic conclusion is that SD experiments are not a competitive probe of the DM dipole moments. By no means does this imply that SD would not be a useful probe of other DM candidates. And indeed, the very purpose of the released code \codename~is to be able to quickly and robustly assess the constraining power of SD experiments to any specific DM model.

Our results come with a few limitations and approximations worth pointing out. First, we have focused on elastic cross sections for which the DM starts thermally coupled, and eventually decouples from the plasma. In particular, we do not consider Coulomb-like interactions with baryons $\sigma_{\chi b}(v) \propto 1/v^4$, which would result from a ``millicharged" DM particle \cite{Dubovsky_01, Dubovsky_04}. We defer a thorough treatment of this case to future work, as it requires self-consistently accounting for bulk relative velocities between baryons and DM \cite{Tseliakhovich_10, Munoz_15, Boddy_18b, Slatyer_18}. Second, we have only considered non-relativistic DM particles, and made a simple approximation for the temperature evolution in the case where DM thermally decouples before it becomes non-relativistic. Improving upon this will require parametrizing the interactions in a Lorentz-invariant fashion, and generalizing the evolution equation for the DM temperature. Last but not least, we assumed that the DM has a Maxwell-Boltzmann velocity distribution, which need not be the case after thermal decoupling, unless the DM efficiently self-interacts \cite{YAH_19}. Our treatment can be generalized to self-consistently account for DM self interactions by solving the Boltzmann equation for the DM velocity distribution, for instance with a Fokker-Planck approximation \cite{YAH_19}; we defer this extension to future work.

In closing, if DM is made entirely or in part of a yet undiscovered particle, robustly identifying its specific nature will likely require combining several experimental, astrophysical and cosmological probes. In this work we have taken the first step towards building an exhaustive framework for the signatures of DM interactions in CMB spectral distortions, which are a promising avenue to probe the nature of DM. The numerical code presented here should provide a valuable addition to the dark matter hunter toolkit.

\section*{Acknowledgements}

I thank Jens Chluba and Vera Gluscevic for useful conversations. Am I grateful to Kimberly Boddy for thorough comments on this manuscript, to Ken Van Tilburg for discussions on the perturbativity limit of dipole moments, and to Tracy Slatyer for clarifications on the effective deposition efficiencies for DM annihilation. I acknowledge support from the National Science Foundation through award number 1820861 and from NASA through grant number 80NSSC20K0532.

\appendix

\section{Analytic solution for DM-photon scattering in the radiation era}\label{app:analytic}

\begin{figure}[h]
\includegraphics[width = \columnwidth]{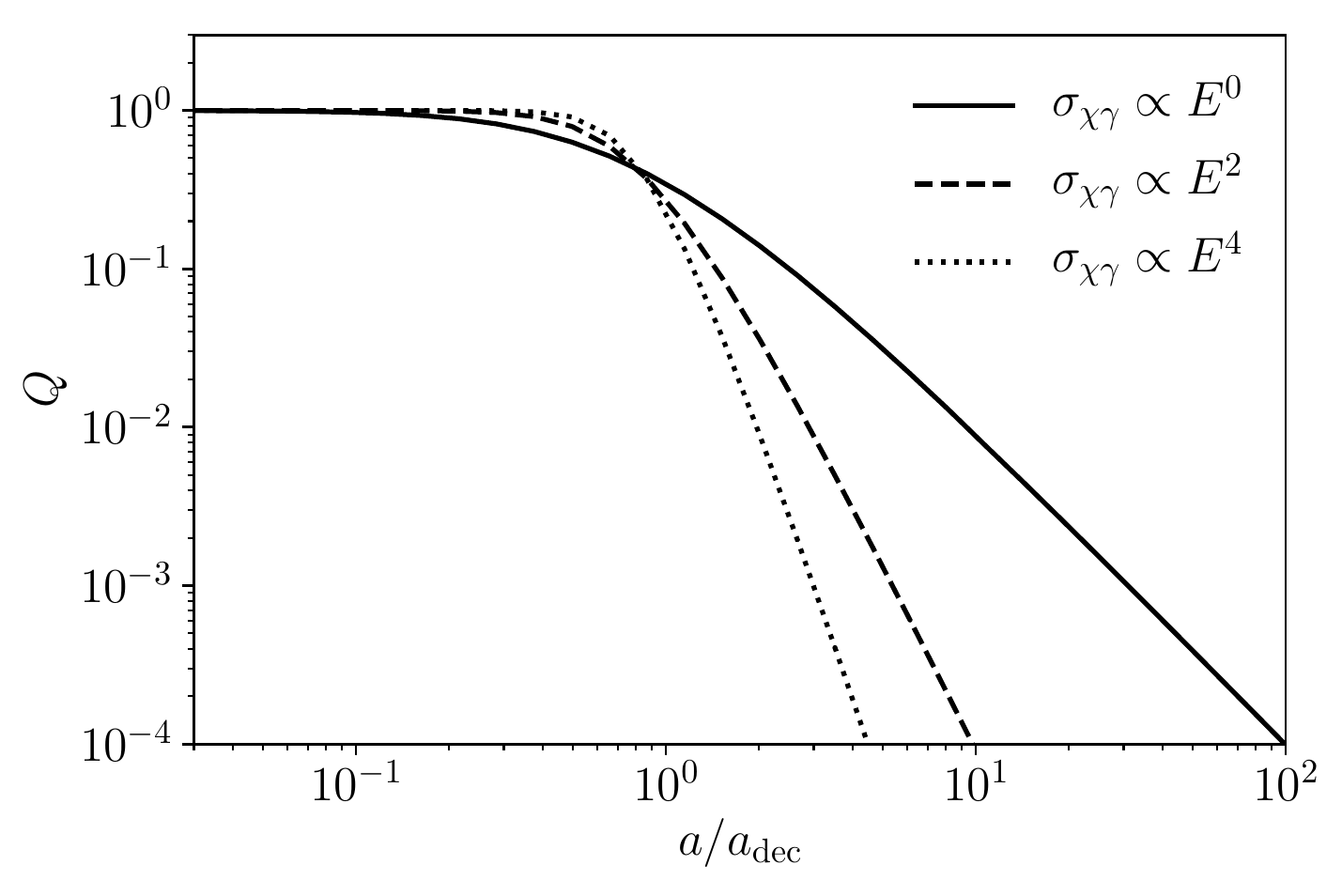}
\caption{Exact dimensionless heating rate for DM-photon scattering in the radiation-dominated era.}
\label{fig:Q_photon}
\end{figure}

Here we consider DM scattering only with photons. During the radiation-dominated era, $H = H_0 \Omega_r^{1/2} a^{-2}$. DM thermal decoupling occurs at a characteristic scale factor $a_{\rm dec}$ given by  \citep{ACK15}
\beq
a_{\rm dec}^{p+2} \equiv \frac83 \frac{d_p \sigma_p \rho_\gamma^0}{m_\chi H_0 \Omega_r^{1/2}}\left(\frac{T_0}{m_\chi}\right)^p. \label{eq:a_p}
\eeq
We can rewrite Eq.~\eqref{eq:Tchi-ODE} in the following self-similar form:
\barr
\frac{d}{dx}(x X) = x^{-(p+2)} \left(1 - X \right), \\
x \equiv a/a_{\rm dec},  \ \ \ X \equiv T_\chi/T_\gamma.
\earr
The solution can be written as an integral:
\beq
X(x) = \int_u^\infty dv ~\rme^{u-v} (u/v)^{\frac{1}{p+2}}, \ \ \ u \equiv \frac{x^{-(p+2)}}{(p+2)}.
\eeq
The dimensionless heating rate is then obtained from $Q = d(xX)/dx$, which can be expressed in terms of the incomplete Gamma function: 
\beq
Q = (p+2) u \times \left(1 - u^{\frac1{p+2}} \rme^u \Gamma\left(\frac{p+1}{p+2}, u\right) \right).
\eeq
For $a \ll a_{\rm dec}$, $Q \rightarrow 1$. For $a \gg a_{\rm dec}$, we have $Q \approx (a_{\rm dec}/a)^{p+2}$. We show $Q(a/a_{\rm dec})$ for a few values of $p$ in Fig.~\ref{fig:Q_photon}. We have checked explicitly that our numerical ODE integrator reproduces the analytic solution for $Q$ within better than $0.1\%$ accuracy.

\bibliography{dm_scat.bib}

\begin{thebibliography}{73}%
\makeatletter
\providecommand \@ifxundefined [1]{%
 \@ifx{#1\undefined}
}%
\providecommand \@ifnum [1]{%
 \ifnum #1\expandafter \@firstoftwo
 \else \expandafter \@secondoftwo
 \fi
}%
\providecommand \@ifx [1]{%
 \ifx #1\expandafter \@firstoftwo
 \else \expandafter \@secondoftwo
 \fi
}%
\providecommand \natexlab [1]{#1}%
\providecommand \enquote  [1]{``#1''}%
\providecommand \bibnamefont  [1]{#1}%
\providecommand \bibfnamefont [1]{#1}%
\providecommand \citenamefont [1]{#1}%
\providecommand \href@noop [0]{\@secondoftwo}%
\providecommand \href [0]{\begingroup \@sanitize@url \@href}%
\providecommand \@href[1]{\@@startlink{#1}\@@href}%
\providecommand \@@href[1]{\endgroup#1\@@endlink}%
\providecommand \@sanitize@url [0]{\catcode `\\12\catcode `\$12\catcode
  `\&12\catcode `\#12\catcode `\^12\catcode `\_12\catcode `\%12\relax}%
\providecommand \@@startlink[1]{}%
\providecommand \@@endlink[0]{}%
\providecommand \url  [0]{\begingroup\@sanitize@url \@url }%
\providecommand \@url [1]{\endgroup\@href {#1}{\urlprefix }}%
\providecommand \urlprefix  [0]{URL }%
\providecommand \Eprint [0]{\href }%
\providecommand \doibase [0]{http://dx.doi.org/}%
\providecommand \selectlanguage [0]{\@gobble}%
\providecommand \bibinfo  [0]{\@secondoftwo}%
\providecommand \bibfield  [0]{\@secondoftwo}%
\providecommand \translation [1]{[#1]}%
\providecommand \BibitemOpen [0]{}%
\providecommand \bibitemStop [0]{}%
\providecommand \bibitemNoStop [0]{.\EOS\space}%
\providecommand \EOS [0]{\spacefactor3000\relax}%
\providecommand \BibitemShut  [1]{\csname bibitem#1\endcsname}%
\let\auto@bib@innerbib\@empty
\bibitem [{\citenamefont {{Zeldovich}}\ and\ \citenamefont
  {{Sunyaev}}(1969)}]{ZS_69}%
  \BibitemOpen
  \bibfield  {author} {\bibinfo {author} {\bibfnamefont {Y.~B.}\ \bibnamefont
  {{Zeldovich}}}\ and\ \bibinfo {author} {\bibfnamefont {R.~A.}\ \bibnamefont
  {{Sunyaev}}},\ }\href {\doibase 10.1007/BF00661821} {\bibfield  {journal}
  {\bibinfo  {journal} {\apss}\ }\textbf {\bibinfo {volume} {4}},\ \bibinfo
  {pages} {301} (\bibinfo {year} {1969})}\BibitemShut {NoStop}%
\bibitem [{\citenamefont {{Sunyaev}}\ and\ \citenamefont
  {{Zeldovich}}(1969)}]{SZ_69}%
  \BibitemOpen
  \bibfield  {author} {\bibinfo {author} {\bibfnamefont {R.~A.}\ \bibnamefont
  {{Sunyaev}}}\ and\ \bibinfo {author} {\bibfnamefont {Y.~B.}\ \bibnamefont
  {{Zeldovich}}},\ }\href {\doibase 10.1038/223721a0} {\bibfield  {journal}
  {\bibinfo  {journal} {\nat}\ }\textbf {\bibinfo {volume} {223}},\ \bibinfo
  {pages} {721} (\bibinfo {year} {1969})}\BibitemShut {NoStop}%
\bibitem [{\citenamefont {{Sunyaev}}\ and\ \citenamefont
  {{Zeldovich}}(1970{\natexlab{a}})}]{SZ_70}%
  \BibitemOpen
  \bibfield  {author} {\bibinfo {author} {\bibfnamefont {R.~A.}\ \bibnamefont
  {{Sunyaev}}}\ and\ \bibinfo {author} {\bibfnamefont {Y.~B.}\ \bibnamefont
  {{Zeldovich}}},\ }\href {\doibase 10.1007/BF00653472} {\bibfield  {journal}
  {\bibinfo  {journal} {\apss}\ }\textbf {\bibinfo {volume} {7}},\ \bibinfo
  {pages} {20} (\bibinfo {year} {1970}{\natexlab{a}})}\BibitemShut {NoStop}%
\bibitem [{\citenamefont {{Sunyaev}}\ and\ \citenamefont
  {{Zeldovich}}(1970{\natexlab{b}})}]{SZ_70b}%
  \BibitemOpen
  \bibfield  {author} {\bibinfo {author} {\bibfnamefont {R.~A.}\ \bibnamefont
  {{Sunyaev}}}\ and\ \bibinfo {author} {\bibfnamefont {Y.~B.}\ \bibnamefont
  {{Zeldovich}}},\ }\href@noop {} {\bibfield  {journal} {\bibinfo  {journal}
  {Comments on Astrophysics and Space Physics}\ }\textbf {\bibinfo {volume}
  {2}},\ \bibinfo {pages} {66} (\bibinfo {year}
  {1970}{\natexlab{b}})}\BibitemShut {NoStop}%
\bibitem [{\citenamefont {{Chluba}}(2016)}]{Chluba_16}%
  \BibitemOpen
  \bibfield  {author} {\bibinfo {author} {\bibfnamefont {J.}~\bibnamefont
  {{Chluba}}},\ }\href {\doibase 10.1093/mnras/stw945} {\bibfield  {journal}
  {\bibinfo  {journal} {\mnras}\ }\textbf {\bibinfo {volume} {460}},\ \bibinfo
  {pages} {227} (\bibinfo {year} {2016})},\ \Eprint
  {http://arxiv.org/abs/1603.02496} {arXiv:1603.02496 [astro-ph.CO]}
  \BibitemShut {NoStop}%
\bibitem [{\citenamefont {{Chluba}}\ and\ \citenamefont
  {{Sunyaev}}(2012)}]{Chluba_12}%
  \BibitemOpen
  \bibfield  {author} {\bibinfo {author} {\bibfnamefont {J.}~\bibnamefont
  {{Chluba}}}\ and\ \bibinfo {author} {\bibfnamefont {R.~A.}\ \bibnamefont
  {{Sunyaev}}},\ }\href {\doibase 10.1111/j.1365-2966.2011.19786.x} {\bibfield
  {journal} {\bibinfo  {journal} {\mnras}\ }\textbf {\bibinfo {volume} {419}},\
  \bibinfo {pages} {1294} (\bibinfo {year} {2012})},\ \Eprint
  {http://arxiv.org/abs/1109.6552} {arXiv:1109.6552} \BibitemShut {NoStop}%
\bibitem [{\citenamefont {{Rubi{\~n}o-Mart{\'{\i}}n}}\ \emph
  {et~al.}(2008)\citenamefont {{Rubi{\~n}o-Mart{\'{\i}}n}}, \citenamefont
  {{Chluba}},\ and\ \citenamefont {{Sunyaev}}}]{Rubino_08}%
  \BibitemOpen
  \bibfield  {author} {\bibinfo {author} {\bibfnamefont {J.~A.}\ \bibnamefont
  {{Rubi{\~n}o-Mart{\'{\i}}n}}}, \bibinfo {author} {\bibfnamefont
  {J.}~\bibnamefont {{Chluba}}}, \ and\ \bibinfo {author} {\bibfnamefont
  {R.~A.}\ \bibnamefont {{Sunyaev}}},\ }\href {\doibase
  10.1051/0004-6361:20078993} {\bibfield  {journal} {\bibinfo  {journal}
  {\aap}\ }\textbf {\bibinfo {volume} {485}},\ \bibinfo {pages} {377} (\bibinfo
  {year} {2008})},\ \Eprint {http://arxiv.org/abs/0711.0594} {arXiv:0711.0594}
  \BibitemShut {NoStop}%
\bibitem [{\citenamefont {{Sunyaev}}\ and\ \citenamefont
  {{Chluba}}(2009)}]{Sunyaev_09}%
  \BibitemOpen
  \bibfield  {author} {\bibinfo {author} {\bibfnamefont {R.~A.}\ \bibnamefont
  {{Sunyaev}}}\ and\ \bibinfo {author} {\bibfnamefont {J.}~\bibnamefont
  {{Chluba}}},\ }\href {\doibase 10.1002/asna.200911237} {\bibfield  {journal}
  {\bibinfo  {journal} {Astronomische Nachrichten}\ }\textbf {\bibinfo {volume}
  {330}},\ \bibinfo {pages} {657} (\bibinfo {year} {2009})},\ \Eprint
  {http://arxiv.org/abs/0908.0435} {arXiv:0908.0435 [astro-ph.CO]} \BibitemShut
  {NoStop}%
\bibitem [{\citenamefont {{Ali-Ha{\"\i}moud}}(2013)}]{YAH_13}%
  \BibitemOpen
  \bibfield  {author} {\bibinfo {author} {\bibfnamefont {Y.}~\bibnamefont
  {{Ali-Ha{\"\i}moud}}},\ }\href {\doibase 10.1103/PhysRevD.87.023526}
  {\bibfield  {journal} {\bibinfo  {journal} {\prd}\ }\textbf {\bibinfo
  {volume} {87}},\ \bibinfo {eid} {023526} (\bibinfo {year} {2013})},\ \Eprint
  {http://arxiv.org/abs/1211.4031} {arXiv:1211.4031 [astro-ph.CO]} \BibitemShut
  {NoStop}%
\bibitem [{\citenamefont {{Chluba}}\ and\ \citenamefont
  {{Ali-Ha{\"\i}moud}}(2016)}]{Chluba_16b}%
  \BibitemOpen
  \bibfield  {author} {\bibinfo {author} {\bibfnamefont {J.}~\bibnamefont
  {{Chluba}}}\ and\ \bibinfo {author} {\bibfnamefont {Y.}~\bibnamefont
  {{Ali-Ha{\"\i}moud}}},\ }\href {\doibase 10.1093/mnras/stv2691} {\bibfield
  {journal} {\bibinfo  {journal} {\mnras}\ }\textbf {\bibinfo {volume} {456}},\
  \bibinfo {pages} {3494} (\bibinfo {year} {2016})},\ \Eprint
  {http://arxiv.org/abs/1510.03877} {arXiv:1510.03877 [astro-ph.CO]}
  \BibitemShut {NoStop}%
\bibitem [{\citenamefont {{Silk}}(1968)}]{Silk_68}%
  \BibitemOpen
  \bibfield  {author} {\bibinfo {author} {\bibfnamefont {J.}~\bibnamefont
  {{Silk}}},\ }\href {\doibase 10.1086/149449} {\bibfield  {journal} {\bibinfo
  {journal} {\apj}\ }\textbf {\bibinfo {volume} {151}},\ \bibinfo {pages} {459}
  (\bibinfo {year} {1968})}\BibitemShut {NoStop}%
\bibitem [{\citenamefont {{Barrow}}\ and\ \citenamefont
  {{Coles}}(1991)}]{Barrow_91}%
  \BibitemOpen
  \bibfield  {author} {\bibinfo {author} {\bibfnamefont {J.~D.}\ \bibnamefont
  {{Barrow}}}\ and\ \bibinfo {author} {\bibfnamefont {P.}~\bibnamefont
  {{Coles}}},\ }\href {\doibase 10.1093/mnras/248.1.52} {\bibfield  {journal}
  {\bibinfo  {journal} {\mnras}\ }\textbf {\bibinfo {volume} {248}},\ \bibinfo
  {pages} {52} (\bibinfo {year} {1991})}\BibitemShut {NoStop}%
\bibitem [{\citenamefont {{Daly}}(1991)}]{Daly_91}%
  \BibitemOpen
  \bibfield  {author} {\bibinfo {author} {\bibfnamefont {R.~A.}\ \bibnamefont
  {{Daly}}},\ }\href {\doibase 10.1086/169866} {\bibfield  {journal} {\bibinfo
  {journal} {\apj}\ }\textbf {\bibinfo {volume} {371}},\ \bibinfo {pages} {14}
  (\bibinfo {year} {1991})}\BibitemShut {NoStop}%
\bibitem [{\citenamefont {{Hu}}\ \emph {et~al.}(1994)\citenamefont {{Hu}},
  \citenamefont {{Scott}},\ and\ \citenamefont {{Silk}}}]{Hu_94}%
  \BibitemOpen
  \bibfield  {author} {\bibinfo {author} {\bibfnamefont {W.}~\bibnamefont
  {{Hu}}}, \bibinfo {author} {\bibfnamefont {D.}~\bibnamefont {{Scott}}}, \
  and\ \bibinfo {author} {\bibfnamefont {J.}~\bibnamefont {{Silk}}},\ }\href
  {\doibase 10.1086/187424} {\bibfield  {journal} {\bibinfo  {journal} {\apjl}\
  }\textbf {\bibinfo {volume} {430}},\ \bibinfo {pages} {L5} (\bibinfo {year}
  {1994})},\ \Eprint {http://arxiv.org/abs/astro-ph/9402045}
  {arXiv:astro-ph/9402045 [astro-ph]} \BibitemShut {NoStop}%
\bibitem [{\citenamefont {{Pajer}}\ and\ \citenamefont
  {{Zaldarriaga}}(2013)}]{Pajer_13}%
  \BibitemOpen
  \bibfield  {author} {\bibinfo {author} {\bibfnamefont {E.}~\bibnamefont
  {{Pajer}}}\ and\ \bibinfo {author} {\bibfnamefont {M.}~\bibnamefont
  {{Zaldarriaga}}},\ }\href {\doibase 10.1088/1475-7516/2013/02/036} {\bibfield
   {journal} {\bibinfo  {journal} {\jcap}\ }\textbf {\bibinfo {volume}
  {2013}},\ \bibinfo {eid} {036} (\bibinfo {year} {2013})},\ \Eprint
  {http://arxiv.org/abs/1206.4479} {arXiv:1206.4479 [astro-ph.CO]} \BibitemShut
  {NoStop}%
\bibitem [{\citenamefont {{Hill}}\ \emph {et~al.}(2015)\citenamefont {{Hill}},
  \citenamefont {{Battaglia}}, \citenamefont {{Chluba}}, \citenamefont
  {{Ferraro}}, \citenamefont {{Schaan}},\ and\ \citenamefont
  {{Spergel}}}]{Hill_15}%
  \BibitemOpen
  \bibfield  {author} {\bibinfo {author} {\bibfnamefont {J.~C.}\ \bibnamefont
  {{Hill}}}, \bibinfo {author} {\bibfnamefont {N.}~\bibnamefont {{Battaglia}}},
  \bibinfo {author} {\bibfnamefont {J.}~\bibnamefont {{Chluba}}}, \bibinfo
  {author} {\bibfnamefont {S.}~\bibnamefont {{Ferraro}}}, \bibinfo {author}
  {\bibfnamefont {E.}~\bibnamefont {{Schaan}}}, \ and\ \bibinfo {author}
  {\bibfnamefont {D.~N.}\ \bibnamefont {{Spergel}}},\ }\href {\doibase
  10.1103/PhysRevLett.115.261301} {\bibfield  {journal} {\bibinfo  {journal}
  {\prl}\ }\textbf {\bibinfo {volume} {115}},\ \bibinfo {eid} {261301}
  (\bibinfo {year} {2015})},\ \Eprint {http://arxiv.org/abs/1507.01583}
  {arXiv:1507.01583 [astro-ph.CO]} \BibitemShut {NoStop}%
\bibitem [{\citenamefont {{Fixsen}}\ \emph {et~al.}(1996)\citenamefont
  {{Fixsen}}, \citenamefont {{Cheng}}, \citenamefont {{Gales}}, \citenamefont
  {{Mather}}, \citenamefont {{Shafer}},\ and\ \citenamefont
  {{Wright}}}]{Fixen_96}%
  \BibitemOpen
  \bibfield  {author} {\bibinfo {author} {\bibfnamefont {D.~J.}\ \bibnamefont
  {{Fixsen}}}, \bibinfo {author} {\bibfnamefont {E.~S.}\ \bibnamefont
  {{Cheng}}}, \bibinfo {author} {\bibfnamefont {J.~M.}\ \bibnamefont
  {{Gales}}}, \bibinfo {author} {\bibfnamefont {J.~C.}\ \bibnamefont
  {{Mather}}}, \bibinfo {author} {\bibfnamefont {R.~A.}\ \bibnamefont
  {{Shafer}}}, \ and\ \bibinfo {author} {\bibfnamefont {E.~L.}\ \bibnamefont
  {{Wright}}},\ }\href {\doibase 10.1086/178173} {\bibfield  {journal}
  {\bibinfo  {journal} {\apj}\ }\textbf {\bibinfo {volume} {473}},\ \bibinfo
  {pages} {576} (\bibinfo {year} {1996})},\ \Eprint
  {http://arxiv.org/abs/astro-ph/9605054} {astro-ph/9605054} \BibitemShut
  {NoStop}%
\bibitem [{\citenamefont {{McDonald}}\ \emph {et~al.}(2001)\citenamefont
  {{McDonald}}, \citenamefont {{Scherrer}},\ and\ \citenamefont
  {{Walker}}}]{McDonald_00}%
  \BibitemOpen
  \bibfield  {author} {\bibinfo {author} {\bibfnamefont {P.}~\bibnamefont
  {{McDonald}}}, \bibinfo {author} {\bibfnamefont {R.~J.}\ \bibnamefont
  {{Scherrer}}}, \ and\ \bibinfo {author} {\bibfnamefont {T.~P.}\ \bibnamefont
  {{Walker}}},\ }\href {\doibase 10.1103/PhysRevD.63.023001} {\bibfield
  {journal} {\bibinfo  {journal} {\prd}\ }\textbf {\bibinfo {volume} {63}},\
  \bibinfo {eid} {023001} (\bibinfo {year} {2001})},\ \Eprint
  {http://arxiv.org/abs/astro-ph/0008134} {arXiv:astro-ph/0008134 [astro-ph]}
  \BibitemShut {NoStop}%
\bibitem [{\citenamefont {{Ali-Ha{\"i}moud}}\ \emph {et~al.}(2015)\citenamefont
  {{Ali-Ha{\"i}moud}}, \citenamefont {{Chluba}},\ and\ \citenamefont
  {{Kamionkowski}}}]{ACK15}%
  \BibitemOpen
  \bibfield  {author} {\bibinfo {author} {\bibfnamefont {Y.}~\bibnamefont
  {{Ali-Ha{\"i}moud}}}, \bibinfo {author} {\bibfnamefont {J.}~\bibnamefont
  {{Chluba}}}, \ and\ \bibinfo {author} {\bibfnamefont {M.}~\bibnamefont
  {{Kamionkowski}}},\ }\href {\doibase 10.1103/PhysRevLett.115.071304}
  {\bibfield  {journal} {\bibinfo  {journal} {Physical Review Letters}\
  }\textbf {\bibinfo {volume} {115}},\ \bibinfo {eid} {071304} (\bibinfo {year}
  {2015})},\ \Eprint {http://arxiv.org/abs/1506.04745} {arXiv:1506.04745}
  \BibitemShut {NoStop}%
\bibitem [{\citenamefont {{Kogut}}\ \emph {et~al.}(2011)\citenamefont
  {{Kogut}}, \citenamefont {{Fixsen}}, \citenamefont {{Chuss}}, \citenamefont
  {{Dotson}}, \citenamefont {{Dwek}}, \citenamefont {{Halpern}}, \citenamefont
  {{Hinshaw}}, \citenamefont {{Meyer}}, \citenamefont {{Moseley}},
  \citenamefont {{Seiffert}}, \citenamefont {{Spergel}},\ and\ \citenamefont
  {{Wollack}}}]{Kogut_11}%
  \BibitemOpen
  \bibfield  {author} {\bibinfo {author} {\bibfnamefont {A.}~\bibnamefont
  {{Kogut}}}, \bibinfo {author} {\bibfnamefont {D.~J.}\ \bibnamefont
  {{Fixsen}}}, \bibinfo {author} {\bibfnamefont {D.~T.}\ \bibnamefont
  {{Chuss}}}, \bibinfo {author} {\bibfnamefont {J.}~\bibnamefont {{Dotson}}},
  \bibinfo {author} {\bibfnamefont {E.}~\bibnamefont {{Dwek}}}, \bibinfo
  {author} {\bibfnamefont {M.}~\bibnamefont {{Halpern}}}, \bibinfo {author}
  {\bibfnamefont {G.~F.}\ \bibnamefont {{Hinshaw}}}, \bibinfo {author}
  {\bibfnamefont {S.~M.}\ \bibnamefont {{Meyer}}}, \bibinfo {author}
  {\bibfnamefont {S.~H.}\ \bibnamefont {{Moseley}}}, \bibinfo {author}
  {\bibfnamefont {M.~D.}\ \bibnamefont {{Seiffert}}}, \bibinfo {author}
  {\bibfnamefont {D.~N.}\ \bibnamefont {{Spergel}}}, \ and\ \bibinfo {author}
  {\bibfnamefont {E.~J.}\ \bibnamefont {{Wollack}}},\ }\href {\doibase
  10.1088/1475-7516/2011/07/025} {\bibfield  {journal} {\bibinfo  {journal}
  {\jcap}\ }\textbf {\bibinfo {volume} {7}},\ \bibinfo {eid} {025} (\bibinfo
  {year} {2011})},\ \Eprint {http://arxiv.org/abs/1105.2044} {arXiv:1105.2044}
  \BibitemShut {NoStop}%
\bibitem [{\citenamefont {{Ali-Ha{\"\i}moud}}(2019)}]{YAH_19}%
  \BibitemOpen
  \bibfield  {author} {\bibinfo {author} {\bibfnamefont {Y.}~\bibnamefont
  {{Ali-Ha{\"\i}moud}}},\ }\href {\doibase 10.1103/PhysRevD.99.023523}
  {\bibfield  {journal} {\bibinfo  {journal} {\prd}\ }\textbf {\bibinfo
  {volume} {99}},\ \bibinfo {eid} {023523} (\bibinfo {year} {2019})},\ \Eprint
  {http://arxiv.org/abs/1811.09903} {arXiv:1811.09903 [astro-ph.CO]}
  \BibitemShut {NoStop}%
\bibitem [{\citenamefont {{Bertschinger}}(2006)}]{Bertschinger_06}%
  \BibitemOpen
  \bibfield  {author} {\bibinfo {author} {\bibfnamefont {E.}~\bibnamefont
  {{Bertschinger}}},\ }\href {\doibase 10.1103/PhysRevD.74.063509} {\bibfield
  {journal} {\bibinfo  {journal} {\prd}\ }\textbf {\bibinfo {volume} {74}},\
  \bibinfo {eid} {063509} (\bibinfo {year} {2006})},\ \Eprint
  {http://arxiv.org/abs/astro-ph/0607319} {astro-ph/0607319} \BibitemShut
  {NoStop}%
\bibitem [{\citenamefont {{Aghanim}}\ \emph {et~al.}(2020)\citenamefont
  {{Aghanim}} \emph {et~al.}}]{Planck_18}%
  \BibitemOpen
  \bibfield  {author} {\bibinfo {author} {\bibfnamefont {N.}~\bibnamefont
  {{Aghanim}}} \emph {et~al.} (\bibinfo {collaboration} {Planck
  collaboration}),\ }\href {\doibase 10.1051/0004-6361/201833910} {\bibfield
  {journal} {\bibinfo  {journal} {\aap}\ }\textbf {\bibinfo {volume} {641}},\
  \bibinfo {eid} {A6} (\bibinfo {year} {2020})},\ \Eprint
  {http://arxiv.org/abs/1807.06209} {arXiv:1807.06209 [astro-ph.CO]}
  \BibitemShut {NoStop}%
\bibitem [{\citenamefont {{Chluba}}(2018)}]{Chluba_18}%
  \BibitemOpen
  \bibfield  {author} {\bibinfo {author} {\bibfnamefont {J.}~\bibnamefont
  {{Chluba}}},\ }\href@noop {} {\bibfield  {journal} {\bibinfo  {journal}
  {arXiv e-prints}\ ,\ \bibinfo {eid} {arXiv:1806.02915}} (\bibinfo {year}
  {2018})},\ \Eprint {http://arxiv.org/abs/1806.02915} {arXiv:1806.02915
  [astro-ph.CO]} \BibitemShut {NoStop}%
\bibitem [{\citenamefont {{Chluba}}(2013{\natexlab{a}})}]{Chluba_13}%
  \BibitemOpen
  \bibfield  {author} {\bibinfo {author} {\bibfnamefont {J.}~\bibnamefont
  {{Chluba}}},\ }\href {\doibase 10.1093/mnras/stt1025} {\bibfield  {journal}
  {\bibinfo  {journal} {\mnras}\ }\textbf {\bibinfo {volume} {434}},\ \bibinfo
  {pages} {352} (\bibinfo {year} {2013}{\natexlab{a}})},\ \Eprint
  {http://arxiv.org/abs/1304.6120} {arXiv:1304.6120} \BibitemShut {NoStop}%
\bibitem [{\citenamefont {{Chluba}}(2015)}]{Chluba_15}%
  \BibitemOpen
  \bibfield  {author} {\bibinfo {author} {\bibfnamefont {J.}~\bibnamefont
  {{Chluba}}},\ }\href {\doibase 10.1093/mnras/stv2243} {\bibfield  {journal}
  {\bibinfo  {journal} {\mnras}\ }\textbf {\bibinfo {volume} {454}},\ \bibinfo
  {pages} {4182} (\bibinfo {year} {2015})},\ \Eprint
  {http://arxiv.org/abs/1506.06582} {arXiv:1506.06582 [astro-ph.CO]}
  \BibitemShut {NoStop}%
\bibitem [{\citenamefont {{Chluba}}(2014)}]{Chluba_14}%
  \BibitemOpen
  \bibfield  {author} {\bibinfo {author} {\bibfnamefont {J.}~\bibnamefont
  {{Chluba}}},\ }\href {\doibase 10.1093/mnras/stu414} {\bibfield  {journal}
  {\bibinfo  {journal} {\mnras}\ }\textbf {\bibinfo {volume} {440}},\ \bibinfo
  {pages} {2544} (\bibinfo {year} {2014})},\ \Eprint
  {http://arxiv.org/abs/1312.6030} {arXiv:1312.6030 [astro-ph.CO]} \BibitemShut
  {NoStop}%
\bibitem [{\citenamefont {{Dvorkin}}\ \emph {et~al.}(2014)\citenamefont
  {{Dvorkin}}, \citenamefont {{Blum}},\ and\ \citenamefont
  {{Kamionkowski}}}]{Dvorkin_14}%
  \BibitemOpen
  \bibfield  {author} {\bibinfo {author} {\bibfnamefont {C.}~\bibnamefont
  {{Dvorkin}}}, \bibinfo {author} {\bibfnamefont {K.}~\bibnamefont {{Blum}}}, \
  and\ \bibinfo {author} {\bibfnamefont {M.}~\bibnamefont {{Kamionkowski}}},\
  }\href {\doibase 10.1103/PhysRevD.89.023519} {\bibfield  {journal} {\bibinfo
  {journal} {\prd}\ }\textbf {\bibinfo {volume} {89}},\ \bibinfo {eid} {023519}
  (\bibinfo {year} {2014})},\ \Eprint {http://arxiv.org/abs/1311.2937}
  {arXiv:1311.2937} \BibitemShut {NoStop}%
\bibitem [{\citenamefont {{Mu{\~n}oz}}\ \emph {et~al.}(2015)\citenamefont
  {{Mu{\~n}oz}}, \citenamefont {{Kovetz}},\ and\ \citenamefont
  {{Ali-Ha{\"i}moud}}}]{Munoz_15}%
  \BibitemOpen
  \bibfield  {author} {\bibinfo {author} {\bibfnamefont {J.~B.}\ \bibnamefont
  {{Mu{\~n}oz}}}, \bibinfo {author} {\bibfnamefont {E.~D.}\ \bibnamefont
  {{Kovetz}}}, \ and\ \bibinfo {author} {\bibfnamefont {Y.}~\bibnamefont
  {{Ali-Ha{\"i}moud}}},\ }\href {\doibase 10.1103/PhysRevD.92.083528}
  {\bibfield  {journal} {\bibinfo  {journal} {\prd}\ }\textbf {\bibinfo
  {volume} {92}},\ \bibinfo {eid} {083528} (\bibinfo {year} {2015})},\ \Eprint
  {http://arxiv.org/abs/1509.00029} {arXiv:1509.00029} \BibitemShut {NoStop}%
\bibitem [{\citenamefont {{Weymann}}(1965)}]{Weymann_65}%
  \BibitemOpen
  \bibfield  {author} {\bibinfo {author} {\bibfnamefont {R.}~\bibnamefont
  {{Weymann}}},\ }\href {\doibase 10.1063/1.1761165} {\bibfield  {journal}
  {\bibinfo  {journal} {Physics of Fluids}\ }\textbf {\bibinfo {volume} {8}},\
  \bibinfo {pages} {2112} (\bibinfo {year} {1965})}\BibitemShut {NoStop}%
\bibitem [{\citenamefont {{Boddy}}\ and\ \citenamefont
  {{Gluscevic}}(2018)}]{Boddy_18}%
  \BibitemOpen
  \bibfield  {author} {\bibinfo {author} {\bibfnamefont {K.~K.}\ \bibnamefont
  {{Boddy}}}\ and\ \bibinfo {author} {\bibfnamefont {V.}~\bibnamefont
  {{Gluscevic}}},\ }\href@noop {} {\bibfield  {journal} {\bibinfo  {journal}
  {ArXiv e-prints}\ } (\bibinfo {year} {2018})},\ \Eprint
  {http://arxiv.org/abs/1801.08609} {arXiv:1801.08609} \BibitemShut {NoStop}%
\bibitem [{\citenamefont {Xu}\ and\ \citenamefont {Farrar}(2021)}]{Xu_21}%
  \BibitemOpen
  \bibfield  {author} {\bibinfo {author} {\bibfnamefont {X.}~\bibnamefont
  {Xu}}\ and\ \bibinfo {author} {\bibfnamefont {G.~R.}\ \bibnamefont
  {Farrar}},\ }\href@noop {} {\  (\bibinfo {year} {2021})},\ \Eprint
  {http://arxiv.org/abs/2101.00142} {arXiv:2101.00142 [hep-ph]} \BibitemShut
  {NoStop}%
\bibitem [{\citenamefont {{Fitzpatrick}}\ \emph {et~al.}(2013)\citenamefont
  {{Fitzpatrick}}, \citenamefont {{Haxton}}, \citenamefont {{Katz}},
  \citenamefont {{Lubbers}},\ and\ \citenamefont {{Xu}}}]{Fitzpatrick_13}%
  \BibitemOpen
  \bibfield  {author} {\bibinfo {author} {\bibfnamefont {A.~L.}\ \bibnamefont
  {{Fitzpatrick}}}, \bibinfo {author} {\bibfnamefont {W.}~\bibnamefont
  {{Haxton}}}, \bibinfo {author} {\bibfnamefont {E.}~\bibnamefont {{Katz}}},
  \bibinfo {author} {\bibfnamefont {N.}~\bibnamefont {{Lubbers}}}, \ and\
  \bibinfo {author} {\bibfnamefont {Y.}~\bibnamefont {{Xu}}},\ }\href {\doibase
  10.1088/1475-7516/2013/02/004} {\bibfield  {journal} {\bibinfo  {journal}
  {\jcap}\ }\textbf {\bibinfo {volume} {2}},\ \bibinfo {eid} {004} (\bibinfo
  {year} {2013})},\ \Eprint {http://arxiv.org/abs/1203.3542} {arXiv:1203.3542
  [hep-ph]} \BibitemShut {NoStop}%
\bibitem [{\citenamefont {{Anand}}\ \emph {et~al.}(2013)\citenamefont
  {{Anand}}, \citenamefont {{Fitzpatrick}},\ and\ \citenamefont
  {{Haxton}}}]{Anand_13}%
  \BibitemOpen
  \bibfield  {author} {\bibinfo {author} {\bibfnamefont {N.}~\bibnamefont
  {{Anand}}}, \bibinfo {author} {\bibfnamefont {A.~L.}\ \bibnamefont
  {{Fitzpatrick}}}, \ and\ \bibinfo {author} {\bibfnamefont {W.~C.}\
  \bibnamefont {{Haxton}}},\ }\href@noop {} {\bibfield  {journal} {\bibinfo
  {journal} {ArXiv e-prints}\ } (\bibinfo {year} {2013})},\ \Eprint
  {http://arxiv.org/abs/1308.6288} {arXiv:1308.6288 [hep-ph]} \BibitemShut
  {NoStop}%
\bibitem [{\citenamefont {{Gluscevic}}\ and\ \citenamefont
  {{Boddy}}(2018)}]{Gluscevic_18}%
  \BibitemOpen
  \bibfield  {author} {\bibinfo {author} {\bibfnamefont {V.}~\bibnamefont
  {{Gluscevic}}}\ and\ \bibinfo {author} {\bibfnamefont {K.~K.}\ \bibnamefont
  {{Boddy}}},\ }\href {\doibase 10.1103/PhysRevLett.121.081301} {\bibfield
  {journal} {\bibinfo  {journal} {\prl}\ }\textbf {\bibinfo {volume} {121}},\
  \bibinfo {eid} {081301} (\bibinfo {year} {2018})},\ \Eprint
  {http://arxiv.org/abs/1712.07133} {arXiv:1712.07133 [astro-ph.CO]}
  \BibitemShut {NoStop}%
\bibitem [{\citenamefont {{Xu}}\ \emph {et~al.}(2018)\citenamefont {{Xu}},
  \citenamefont {{Dvorkin}},\ and\ \citenamefont {{Chael}}}]{Xu_18}%
  \BibitemOpen
  \bibfield  {author} {\bibinfo {author} {\bibfnamefont {W.~L.}\ \bibnamefont
  {{Xu}}}, \bibinfo {author} {\bibfnamefont {C.}~\bibnamefont {{Dvorkin}}}, \
  and\ \bibinfo {author} {\bibfnamefont {A.}~\bibnamefont {{Chael}}},\ }\href
  {\doibase 10.1103/PhysRevD.97.103530} {\bibfield  {journal} {\bibinfo
  {journal} {\prd}\ }\textbf {\bibinfo {volume} {97}},\ \bibinfo {eid} {103530}
  (\bibinfo {year} {2018})},\ \Eprint {http://arxiv.org/abs/1802.06788}
  {arXiv:1802.06788} \BibitemShut {NoStop}%
\bibitem [{\citenamefont {{Maamari}}\ \emph {et~al.}(2020)\citenamefont
  {{Maamari}}, \citenamefont {{Gluscevic}}, \citenamefont {{Boddy}},
  \citenamefont {{Nadler}},\ and\ \citenamefont {{Wechsler}}}]{Maamari_20}%
  \BibitemOpen
  \bibfield  {author} {\bibinfo {author} {\bibfnamefont {K.}~\bibnamefont
  {{Maamari}}}, \bibinfo {author} {\bibfnamefont {V.}~\bibnamefont
  {{Gluscevic}}}, \bibinfo {author} {\bibfnamefont {K.~K.}\ \bibnamefont
  {{Boddy}}}, \bibinfo {author} {\bibfnamefont {E.~O.}\ \bibnamefont
  {{Nadler}}}, \ and\ \bibinfo {author} {\bibfnamefont {R.~H.}\ \bibnamefont
  {{Wechsler}}},\ }\href@noop {} {\bibfield  {journal} {\bibinfo  {journal}
  {arXiv e-prints}\ ,\ \bibinfo {eid} {arXiv:2010.02936}} (\bibinfo {year}
  {2020})},\ \Eprint {http://arxiv.org/abs/2010.02936} {arXiv:2010.02936
  [astro-ph.CO]} \BibitemShut {NoStop}%
\bibitem [{\citenamefont {{Wadekar}}\ and\ \citenamefont
  {{Farrar}}(2019)}]{Wadekar_19}%
  \BibitemOpen
  \bibfield  {author} {\bibinfo {author} {\bibfnamefont {D.}~\bibnamefont
  {{Wadekar}}}\ and\ \bibinfo {author} {\bibfnamefont {G.~R.}\ \bibnamefont
  {{Farrar}}},\ }\href@noop {} {\bibfield  {journal} {\bibinfo  {journal}
  {arXiv e-prints}\ ,\ \bibinfo {eid} {arXiv:1903.12190}} (\bibinfo {year}
  {2019})},\ \Eprint {http://arxiv.org/abs/1903.12190} {arXiv:1903.12190
  [hep-ph]} \BibitemShut {NoStop}%
\bibitem [{\citenamefont {{Boddy}}\ \emph {et~al.}(2018)\citenamefont
  {{Boddy}}, \citenamefont {{Gluscevic}}, \citenamefont {{Poulin}},
  \citenamefont {{Kovetz}}, \citenamefont {{Kamionkowski}},\ and\ \citenamefont
  {{Barkana}}}]{Boddy_18b}%
  \BibitemOpen
  \bibfield  {author} {\bibinfo {author} {\bibfnamefont {K.~K.}\ \bibnamefont
  {{Boddy}}}, \bibinfo {author} {\bibfnamefont {V.}~\bibnamefont
  {{Gluscevic}}}, \bibinfo {author} {\bibfnamefont {V.}~\bibnamefont
  {{Poulin}}}, \bibinfo {author} {\bibfnamefont {E.~D.}\ \bibnamefont
  {{Kovetz}}}, \bibinfo {author} {\bibfnamefont {M.}~\bibnamefont
  {{Kamionkowski}}}, \ and\ \bibinfo {author} {\bibfnamefont {R.}~\bibnamefont
  {{Barkana}}},\ }\href {\doibase 10.1103/PhysRevD.98.123506} {\bibfield
  {journal} {\bibinfo  {journal} {\prd}\ }\textbf {\bibinfo {volume} {98}},\
  \bibinfo {eid} {123506} (\bibinfo {year} {2018})},\ \Eprint
  {http://arxiv.org/abs/1808.00001} {arXiv:1808.00001 [astro-ph.CO]}
  \BibitemShut {NoStop}%
\bibitem [{\citenamefont {{Cappiello}}\ \emph {et~al.}(2018)\citenamefont
  {{Cappiello}}, \citenamefont {{Ng}},\ and\ \citenamefont
  {{Beacom}}}]{Cappiello_18}%
  \BibitemOpen
  \bibfield  {author} {\bibinfo {author} {\bibfnamefont {C.~V.}\ \bibnamefont
  {{Cappiello}}}, \bibinfo {author} {\bibfnamefont {K.~C.~Y.}\ \bibnamefont
  {{Ng}}}, \ and\ \bibinfo {author} {\bibfnamefont {J.~F.}\ \bibnamefont
  {{Beacom}}},\ }\href@noop {} {\bibfield  {journal} {\bibinfo  {journal}
  {ArXiv e-prints}\ } (\bibinfo {year} {2018})},\ \Eprint
  {http://arxiv.org/abs/1810.07705} {arXiv:1810.07705 [hep-ph]} \BibitemShut
  {NoStop}%
\bibitem [{\citenamefont {{Abramoff}}\ \emph {et~al.}(2019)\citenamefont
  {{Abramoff}}, \citenamefont {{Barak}}, \citenamefont {{Bloch}}, \citenamefont
  {{Chaplinsky}}, \citenamefont {{Crisler}}, \citenamefont {{Dawa}},
  \citenamefont {{Drlica-Wagner}}, \citenamefont {{Essig}}, \citenamefont
  {{Estrada}}, \citenamefont {{Etzion}}, \citenamefont {{Fernandez}},
  \citenamefont {{Gift}}, \citenamefont {{Sofo-Haro}}, \citenamefont
  {{Taenzer}}, \citenamefont {{Tiffenberg}}, \citenamefont {{Volansky}},
  \citenamefont {{Yu}},\ and\ \citenamefont {{Sensei
  Collaboration}}}]{Abramoff_19}%
  \BibitemOpen
  \bibfield  {author} {\bibinfo {author} {\bibfnamefont {O.}~\bibnamefont
  {{Abramoff}}}, \bibinfo {author} {\bibfnamefont {L.}~\bibnamefont {{Barak}}},
  \bibinfo {author} {\bibfnamefont {I.~M.}\ \bibnamefont {{Bloch}}}, \bibinfo
  {author} {\bibfnamefont {L.}~\bibnamefont {{Chaplinsky}}}, \bibinfo {author}
  {\bibfnamefont {M.}~\bibnamefont {{Crisler}}}, \bibinfo {author}
  {\bibnamefont {{Dawa}}}, \bibinfo {author} {\bibfnamefont {A.}~\bibnamefont
  {{Drlica-Wagner}}}, \bibinfo {author} {\bibfnamefont {R.}~\bibnamefont
  {{Essig}}}, \bibinfo {author} {\bibfnamefont {J.}~\bibnamefont {{Estrada}}},
  \bibinfo {author} {\bibfnamefont {E.}~\bibnamefont {{Etzion}}}, \bibinfo
  {author} {\bibfnamefont {G.}~\bibnamefont {{Fernandez}}}, \bibinfo {author}
  {\bibfnamefont {D.}~\bibnamefont {{Gift}}}, \bibinfo {author} {\bibfnamefont
  {M.}~\bibnamefont {{Sofo-Haro}}}, \bibinfo {author} {\bibfnamefont
  {J.}~\bibnamefont {{Taenzer}}}, \bibinfo {author} {\bibfnamefont
  {J.}~\bibnamefont {{Tiffenberg}}}, \bibinfo {author} {\bibfnamefont
  {T.}~\bibnamefont {{Volansky}}}, \bibinfo {author} {\bibfnamefont {T.-T.}\
  \bibnamefont {{Yu}}}, \ and\ \bibinfo {author} {\bibnamefont {{Sensei
  Collaboration}}},\ }\href {\doibase 10.1103/PhysRevLett.122.161801}
  {\bibfield  {journal} {\bibinfo  {journal} {\prl}\ }\textbf {\bibinfo
  {volume} {122}},\ \bibinfo {eid} {161801} (\bibinfo {year} {2019})},\ \Eprint
  {http://arxiv.org/abs/1901.10478} {arXiv:1901.10478 [hep-ex]} \BibitemShut
  {NoStop}%
\bibitem [{\citenamefont {{Essig}}\ \emph {et~al.}(2012)\citenamefont
  {{Essig}}, \citenamefont {{Manalaysay}}, \citenamefont {{Mardon}},
  \citenamefont {{Sorensen}},\ and\ \citenamefont {{Volansky}}}]{Essig_12}%
  \BibitemOpen
  \bibfield  {author} {\bibinfo {author} {\bibfnamefont {R.}~\bibnamefont
  {{Essig}}}, \bibinfo {author} {\bibfnamefont {A.}~\bibnamefont
  {{Manalaysay}}}, \bibinfo {author} {\bibfnamefont {J.}~\bibnamefont
  {{Mardon}}}, \bibinfo {author} {\bibfnamefont {P.}~\bibnamefont
  {{Sorensen}}}, \ and\ \bibinfo {author} {\bibfnamefont {T.}~\bibnamefont
  {{Volansky}}},\ }\href {\doibase 10.1103/PhysRevLett.109.021301} {\bibfield
  {journal} {\bibinfo  {journal} {\prl}\ }\textbf {\bibinfo {volume} {109}},\
  \bibinfo {eid} {021301} (\bibinfo {year} {2012})},\ \Eprint
  {http://arxiv.org/abs/1206.2644} {arXiv:1206.2644 [astro-ph.CO]} \BibitemShut
  {NoStop}%
\bibitem [{\citenamefont {{Bernabei}}\ \emph {et~al.}(2008)\citenamefont
  {{Bernabei}}, \citenamefont {{Belli}}, \citenamefont {{Montecchia}},
  \citenamefont {{Nozzoli}}, \citenamefont {{Cappella}}, \citenamefont
  {{Incicchitti}}, \citenamefont {{Prosperi}}, \citenamefont {{Cerulli}},
  \citenamefont {{Dai}}, \citenamefont {{He}}, \citenamefont {{Kuang}},
  \citenamefont {{Ma}}, \citenamefont {{Ma}}, \citenamefont {{Sheng}},
  \citenamefont {{Ye}}, \citenamefont {{Wang}},\ and\ \citenamefont
  {{Zhang}}}]{Bernabei_08}%
  \BibitemOpen
  \bibfield  {author} {\bibinfo {author} {\bibfnamefont {R.}~\bibnamefont
  {{Bernabei}}}, \bibinfo {author} {\bibfnamefont {P.}~\bibnamefont {{Belli}}},
  \bibinfo {author} {\bibfnamefont {F.}~\bibnamefont {{Montecchia}}}, \bibinfo
  {author} {\bibfnamefont {F.}~\bibnamefont {{Nozzoli}}}, \bibinfo {author}
  {\bibfnamefont {F.}~\bibnamefont {{Cappella}}}, \bibinfo {author}
  {\bibfnamefont {A.}~\bibnamefont {{Incicchitti}}}, \bibinfo {author}
  {\bibfnamefont {D.}~\bibnamefont {{Prosperi}}}, \bibinfo {author}
  {\bibfnamefont {R.}~\bibnamefont {{Cerulli}}}, \bibinfo {author}
  {\bibfnamefont {C.~J.}\ \bibnamefont {{Dai}}}, \bibinfo {author}
  {\bibfnamefont {H.~L.}\ \bibnamefont {{He}}}, \bibinfo {author}
  {\bibfnamefont {H.~H.}\ \bibnamefont {{Kuang}}}, \bibinfo {author}
  {\bibfnamefont {J.~M.}\ \bibnamefont {{Ma}}}, \bibinfo {author}
  {\bibfnamefont {X.~H.}\ \bibnamefont {{Ma}}}, \bibinfo {author}
  {\bibfnamefont {X.~D.}\ \bibnamefont {{Sheng}}}, \bibinfo {author}
  {\bibfnamefont {Z.~P.}\ \bibnamefont {{Ye}}}, \bibinfo {author}
  {\bibfnamefont {R.~G.}\ \bibnamefont {{Wang}}}, \ and\ \bibinfo {author}
  {\bibfnamefont {Y.~J.}\ \bibnamefont {{Zhang}}},\ }\href {\doibase
  10.1103/PhysRevD.77.023506} {\bibfield  {journal} {\bibinfo  {journal}
  {\prd}\ }\textbf {\bibinfo {volume} {77}},\ \bibinfo {eid} {023506} (\bibinfo
  {year} {2008})},\ \Eprint {http://arxiv.org/abs/0712.0562} {arXiv:0712.0562
  [astro-ph]} \BibitemShut {NoStop}%
\bibitem [{\citenamefont {{Essig}}\ \emph {et~al.}(2017)\citenamefont
  {{Essig}}, \citenamefont {{Volansky}},\ and\ \citenamefont
  {{Yu}}}]{Essig_17}%
  \BibitemOpen
  \bibfield  {author} {\bibinfo {author} {\bibfnamefont {R.}~\bibnamefont
  {{Essig}}}, \bibinfo {author} {\bibfnamefont {T.}~\bibnamefont {{Volansky}}},
  \ and\ \bibinfo {author} {\bibfnamefont {T.-T.}\ \bibnamefont {{Yu}}},\
  }\href {\doibase 10.1103/PhysRevD.96.043017} {\bibfield  {journal} {\bibinfo
  {journal} {\prd}\ }\textbf {\bibinfo {volume} {96}},\ \bibinfo {eid} {043017}
  (\bibinfo {year} {2017})},\ \Eprint {http://arxiv.org/abs/1703.00910}
  {arXiv:1703.00910 [hep-ph]} \BibitemShut {NoStop}%
\bibitem [{\citenamefont {{Dubovsky}}\ and\ \citenamefont
  {{Gorbunov}}(2001)}]{Dubovsky_01}%
  \BibitemOpen
  \bibfield  {author} {\bibinfo {author} {\bibfnamefont {S.~L.}\ \bibnamefont
  {{Dubovsky}}}\ and\ \bibinfo {author} {\bibfnamefont {D.~S.}\ \bibnamefont
  {{Gorbunov}}},\ }\href {\doibase 10.1103/PhysRevD.64.123503} {\bibfield
  {journal} {\bibinfo  {journal} {\prd}\ }\textbf {\bibinfo {volume} {64}},\
  \bibinfo {pages} {123503} (\bibinfo {year} {2001})},\ \Eprint
  {http://arxiv.org/abs/astro-ph/0103122} {astro-ph/0103122} \BibitemShut
  {NoStop}%
\bibitem [{\citenamefont {{Dubovsky}}\ \emph {et~al.}(2004)\citenamefont
  {{Dubovsky}}, \citenamefont {{Gorbunov}},\ and\ \citenamefont
  {{Rubtsov}}}]{Dubovsky_04}%
  \BibitemOpen
  \bibfield  {author} {\bibinfo {author} {\bibfnamefont {S.~L.}\ \bibnamefont
  {{Dubovsky}}}, \bibinfo {author} {\bibfnamefont {D.~S.}\ \bibnamefont
  {{Gorbunov}}}, \ and\ \bibinfo {author} {\bibfnamefont {G.~I.}\ \bibnamefont
  {{Rubtsov}}},\ }\href {\doibase 10.1134/1.1675909} {\bibfield  {journal}
  {\bibinfo  {journal} {Soviet Journal of Experimental and Theoretical Physics
  Letters}\ }\textbf {\bibinfo {volume} {79}},\ \bibinfo {pages} {1} (\bibinfo
  {year} {2004})},\ \Eprint {http://arxiv.org/abs/hep-ph/0311189}
  {hep-ph/0311189} \BibitemShut {NoStop}%
\bibitem [{\citenamefont {{Sigurdson}}\ \emph {et~al.}(2004)\citenamefont
  {{Sigurdson}}, \citenamefont {{Doran}}, \citenamefont {{Kurylov}},
  \citenamefont {{Caldwell}},\ and\ \citenamefont
  {{Kamionkowski}}}]{Sigurdson_04}%
  \BibitemOpen
  \bibfield  {author} {\bibinfo {author} {\bibfnamefont {K.}~\bibnamefont
  {{Sigurdson}}}, \bibinfo {author} {\bibfnamefont {M.}~\bibnamefont
  {{Doran}}}, \bibinfo {author} {\bibfnamefont {A.}~\bibnamefont {{Kurylov}}},
  \bibinfo {author} {\bibfnamefont {R.~R.}\ \bibnamefont {{Caldwell}}}, \ and\
  \bibinfo {author} {\bibfnamefont {M.}~\bibnamefont {{Kamionkowski}}},\ }\href
  {\doibase 10.1103/PhysRevD.70.083501} {\bibfield  {journal} {\bibinfo
  {journal} {\prd}\ }\textbf {\bibinfo {volume} {70}},\ \bibinfo {eid} {083501}
  (\bibinfo {year} {2004})},\ \Eprint {http://arxiv.org/abs/astro-ph/0406355}
  {astro-ph/0406355} \BibitemShut {NoStop}%
\bibitem [{\citenamefont {{Weiner}}\ and\ \citenamefont
  {{Yavin}}(2012)}]{Weiner_12}%
  \BibitemOpen
  \bibfield  {author} {\bibinfo {author} {\bibfnamefont {N.}~\bibnamefont
  {{Weiner}}}\ and\ \bibinfo {author} {\bibfnamefont {I.}~\bibnamefont
  {{Yavin}}},\ }\href {\doibase 10.1103/PhysRevD.86.075021} {\bibfield
  {journal} {\bibinfo  {journal} {\prd}\ }\textbf {\bibinfo {volume} {86}},\
  \bibinfo {eid} {075021} (\bibinfo {year} {2012})},\ \Eprint
  {http://arxiv.org/abs/1206.2910} {arXiv:1206.2910 [hep-ph]} \BibitemShut
  {NoStop}%
\bibitem [{\citenamefont {{Wilkinson}}\ \emph {et~al.}(2014)\citenamefont
  {{Wilkinson}}, \citenamefont {{Lesgourgues}},\ and\ \citenamefont
  {{B{\oe}hm}}}]{Wilkinson_14a}%
  \BibitemOpen
  \bibfield  {author} {\bibinfo {author} {\bibfnamefont {R.~J.}\ \bibnamefont
  {{Wilkinson}}}, \bibinfo {author} {\bibfnamefont {J.}~\bibnamefont
  {{Lesgourgues}}}, \ and\ \bibinfo {author} {\bibfnamefont {C.}~\bibnamefont
  {{B{\oe}hm}}},\ }\href {\doibase 10.1088/1475-7516/2014/04/026} {\bibfield
  {journal} {\bibinfo  {journal} {\jcap}\ }\textbf {\bibinfo {volume} {4}},\
  \bibinfo {eid} {026} (\bibinfo {year} {2014})},\ \Eprint
  {http://arxiv.org/abs/1309.7588} {arXiv:1309.7588} \BibitemShut {NoStop}%
\bibitem [{\citenamefont {{Stadler}}\ and\ \citenamefont
  {{B{\oe}hm}}(2018)}]{Stadler_18}%
  \BibitemOpen
  \bibfield  {author} {\bibinfo {author} {\bibfnamefont {J.}~\bibnamefont
  {{Stadler}}}\ and\ \bibinfo {author} {\bibfnamefont {C.}~\bibnamefont
  {{B{\oe}hm}}},\ }\href {\doibase 10.1088/1475-7516/2018/10/009} {\bibfield
  {journal} {\bibinfo  {journal} {\jcap}\ }\textbf {\bibinfo {volume} {2018}},\
  \bibinfo {eid} {009} (\bibinfo {year} {2018})},\ \Eprint
  {http://arxiv.org/abs/1802.06589} {arXiv:1802.06589 [astro-ph.CO]}
  \BibitemShut {NoStop}%
\bibitem [{\citenamefont {{Becker}}\ \emph {et~al.}(2020)\citenamefont
  {{Becker}}, \citenamefont {{Hooper}}, \citenamefont {{Kahlhoefer}},
  \citenamefont {{Lesgourgues}},\ and\ \citenamefont
  {{Sch{\"o}neberg}}}]{Becker_20}%
  \BibitemOpen
  \bibfield  {author} {\bibinfo {author} {\bibfnamefont {N.}~\bibnamefont
  {{Becker}}}, \bibinfo {author} {\bibfnamefont {D.~C.}\ \bibnamefont
  {{Hooper}}}, \bibinfo {author} {\bibfnamefont {F.}~\bibnamefont
  {{Kahlhoefer}}}, \bibinfo {author} {\bibfnamefont {J.}~\bibnamefont
  {{Lesgourgues}}}, \ and\ \bibinfo {author} {\bibfnamefont {N.}~\bibnamefont
  {{Sch{\"o}neberg}}},\ }\href@noop {} {\bibfield  {journal} {\bibinfo
  {journal} {arXiv e-prints}\ ,\ \bibinfo {eid} {arXiv:2010.04074}} (\bibinfo
  {year} {2020})},\ \Eprint {http://arxiv.org/abs/2010.04074} {arXiv:2010.04074
  [astro-ph.CO]} \BibitemShut {NoStop}%
\bibitem [{\citenamefont {{B{\oe}hm}}\ \emph {et~al.}(2014)\citenamefont
  {{B{\oe}hm}}, \citenamefont {{Schewtschenko}}, \citenamefont {{Wilkinson}},
  \citenamefont {{Baugh}},\ and\ \citenamefont {{Pascoli}}}]{Boehm_14}%
  \BibitemOpen
  \bibfield  {author} {\bibinfo {author} {\bibfnamefont {C.}~\bibnamefont
  {{B{\oe}hm}}}, \bibinfo {author} {\bibfnamefont {J.~A.}\ \bibnamefont
  {{Schewtschenko}}}, \bibinfo {author} {\bibfnamefont {R.~J.}\ \bibnamefont
  {{Wilkinson}}}, \bibinfo {author} {\bibfnamefont {C.~M.}\ \bibnamefont
  {{Baugh}}}, \ and\ \bibinfo {author} {\bibfnamefont {S.}~\bibnamefont
  {{Pascoli}}},\ }\href {\doibase 10.1093/mnrasl/slu115} {\bibfield  {journal}
  {\bibinfo  {journal} {\mnras}\ }\textbf {\bibinfo {volume} {445}},\ \bibinfo
  {pages} {L31} (\bibinfo {year} {2014})},\ \Eprint
  {http://arxiv.org/abs/1404.7012} {arXiv:1404.7012} \BibitemShut {NoStop}%
\bibitem [{\citenamefont {{Schewtschenko}}\ \emph {et~al.}(2016)\citenamefont
  {{Schewtschenko}}, \citenamefont {{Baugh}}, \citenamefont {{Wilkinson}},
  \citenamefont {{B{\oe}hm}}, \citenamefont {{Pascoli}},\ and\ \citenamefont
  {{Sawala}}}]{Schewtschenko_16}%
  \BibitemOpen
  \bibfield  {author} {\bibinfo {author} {\bibfnamefont {J.~A.}\ \bibnamefont
  {{Schewtschenko}}}, \bibinfo {author} {\bibfnamefont {C.~M.}\ \bibnamefont
  {{Baugh}}}, \bibinfo {author} {\bibfnamefont {R.~J.}\ \bibnamefont
  {{Wilkinson}}}, \bibinfo {author} {\bibfnamefont {C.}~\bibnamefont
  {{B{\oe}hm}}}, \bibinfo {author} {\bibfnamefont {S.}~\bibnamefont
  {{Pascoli}}}, \ and\ \bibinfo {author} {\bibfnamefont {T.}~\bibnamefont
  {{Sawala}}},\ }\href {\doibase 10.1093/mnras/stw1078} {\bibfield  {journal}
  {\bibinfo  {journal} {\mnras}\ }\textbf {\bibinfo {volume} {461}},\ \bibinfo
  {pages} {2282} (\bibinfo {year} {2016})},\ \Eprint
  {http://arxiv.org/abs/1512.06774} {arXiv:1512.06774 [astro-ph.CO]}
  \BibitemShut {NoStop}%
\bibitem [{\citenamefont {{Fortin}}\ and\ \citenamefont
  {{Tait}}(2012)}]{Fortin_12}%
  \BibitemOpen
  \bibfield  {author} {\bibinfo {author} {\bibfnamefont {J.-F.}\ \bibnamefont
  {{Fortin}}}\ and\ \bibinfo {author} {\bibfnamefont {T.~M.~P.}\ \bibnamefont
  {{Tait}}},\ }\href {\doibase 10.1103/PhysRevD.85.063506} {\bibfield
  {journal} {\bibinfo  {journal} {\prd}\ }\textbf {\bibinfo {volume} {85}},\
  \bibinfo {eid} {063506} (\bibinfo {year} {2012})},\ \Eprint
  {http://arxiv.org/abs/1103.3289} {arXiv:1103.3289 [hep-ph]} \BibitemShut
  {NoStop}%
\bibitem [{\citenamefont {{Graham}}\ \emph {et~al.}(2012)\citenamefont
  {{Graham}}, \citenamefont {{Kaplan}}, \citenamefont {{Rajendran}},\ and\
  \citenamefont {{Walters}}}]{Graham_12}%
  \BibitemOpen
  \bibfield  {author} {\bibinfo {author} {\bibfnamefont {P.~W.}\ \bibnamefont
  {{Graham}}}, \bibinfo {author} {\bibfnamefont {D.~E.}\ \bibnamefont
  {{Kaplan}}}, \bibinfo {author} {\bibfnamefont {S.}~\bibnamefont
  {{Rajendran}}}, \ and\ \bibinfo {author} {\bibfnamefont {M.~T.}\ \bibnamefont
  {{Walters}}},\ }\href {\doibase 10.1016/j.dark.2012.09.001} {\bibfield
  {journal} {\bibinfo  {journal} {Physics of the Dark Universe}\ }\textbf
  {\bibinfo {volume} {1}},\ \bibinfo {pages} {32} (\bibinfo {year} {2012})},\
  \Eprint {http://arxiv.org/abs/1203.2531} {arXiv:1203.2531 [hep-ph]}
  \BibitemShut {NoStop}%
\bibitem [{\citenamefont {{Hyeok Chang}}\ \emph {et~al.}(2019)\citenamefont
  {{Hyeok Chang}}, \citenamefont {{Essig}},\ and\ \citenamefont
  {{Reinert}}}]{Chang_19}%
  \BibitemOpen
  \bibfield  {author} {\bibinfo {author} {\bibfnamefont {J.}~\bibnamefont
  {{Hyeok Chang}}}, \bibinfo {author} {\bibfnamefont {R.}~\bibnamefont
  {{Essig}}}, \ and\ \bibinfo {author} {\bibfnamefont {A.}~\bibnamefont
  {{Reinert}}},\ }\href@noop {} {\bibfield  {journal} {\bibinfo  {journal}
  {arXiv e-prints}\ ,\ \bibinfo {eid} {arXiv:1911.03389}} (\bibinfo {year}
  {2019})},\ \Eprint {http://arxiv.org/abs/1911.03389} {arXiv:1911.03389
  [hep-ph]} \BibitemShut {NoStop}%
\bibitem [{\citenamefont {{Gell-Mann}}\ and\ \citenamefont
  {{Goldberger}}(1954)}]{Gell-Mann_54}%
  \BibitemOpen
  \bibfield  {author} {\bibinfo {author} {\bibfnamefont {M.}~\bibnamefont
  {{Gell-Mann}}}\ and\ \bibinfo {author} {\bibfnamefont {M.~L.}\ \bibnamefont
  {{Goldberger}}},\ }\href {\doibase 10.1103/PhysRev.96.1433} {\bibfield
  {journal} {\bibinfo  {journal} {Physical Review}\ }\textbf {\bibinfo {volume}
  {96}},\ \bibinfo {pages} {1433} (\bibinfo {year} {1954})}\BibitemShut
  {NoStop}%
\bibitem [{\citenamefont {{Sigurdson}}\ \emph {et~al.}(2006)\citenamefont
  {{Sigurdson}}, \citenamefont {{Doran}}, \citenamefont {{Kurylov}},
  \citenamefont {{Caldwell}},\ and\ \citenamefont
  {{Kamionkowski}}}]{Sigurdson_04_erratum}%
  \BibitemOpen
  \bibfield  {author} {\bibinfo {author} {\bibfnamefont {K.}~\bibnamefont
  {{Sigurdson}}}, \bibinfo {author} {\bibfnamefont {M.}~\bibnamefont
  {{Doran}}}, \bibinfo {author} {\bibfnamefont {A.}~\bibnamefont {{Kurylov}}},
  \bibinfo {author} {\bibfnamefont {R.~R.}\ \bibnamefont {{Caldwell}}}, \ and\
  \bibinfo {author} {\bibfnamefont {M.}~\bibnamefont {{Kamionkowski}}},\ }\href
  {\doibase 10.1103/PhysRevD.73.089903} {\bibfield  {journal} {\bibinfo
  {journal} {\prd}\ }\textbf {\bibinfo {volume} {73}},\ \bibinfo {eid} {089903}
  (\bibinfo {year} {2006})}\BibitemShut {NoStop}%
\bibitem [{\citenamefont {{Del Nobile}}\ \emph {et~al.}(2012)\citenamefont
  {{Del Nobile}}, \citenamefont {{Kouvaris}}, \citenamefont {{Panci}},
  \citenamefont {{Sannino}},\ and\ \citenamefont
  {{Virkaj{\"a}rvi}}}]{DelNobile_12}%
  \BibitemOpen
  \bibfield  {author} {\bibinfo {author} {\bibfnamefont {E.}~\bibnamefont {{Del
  Nobile}}}, \bibinfo {author} {\bibfnamefont {C.}~\bibnamefont {{Kouvaris}}},
  \bibinfo {author} {\bibfnamefont {P.}~\bibnamefont {{Panci}}}, \bibinfo
  {author} {\bibfnamefont {F.}~\bibnamefont {{Sannino}}}, \ and\ \bibinfo
  {author} {\bibfnamefont {J.}~\bibnamefont {{Virkaj{\"a}rvi}}},\ }\href
  {\doibase 10.1088/1475-7516/2012/08/010} {\bibfield  {journal} {\bibinfo
  {journal} {\jcap}\ }\textbf {\bibinfo {volume} {2012}},\ \bibinfo {eid} {010}
  (\bibinfo {year} {2012})},\ \Eprint {http://arxiv.org/abs/1203.6652}
  {arXiv:1203.6652 [hep-ph]} \BibitemShut {NoStop}%
\bibitem [{\citenamefont {{Mass{\'o}}}\ \emph {et~al.}(2009)\citenamefont
  {{Mass{\'o}}}, \citenamefont {{Mohanty}},\ and\ \citenamefont
  {{Rao}}}]{Masso_09}%
  \BibitemOpen
  \bibfield  {author} {\bibinfo {author} {\bibfnamefont {E.}~\bibnamefont
  {{Mass{\'o}}}}, \bibinfo {author} {\bibfnamefont {S.}~\bibnamefont
  {{Mohanty}}}, \ and\ \bibinfo {author} {\bibfnamefont {S.}~\bibnamefont
  {{Rao}}},\ }\href {\doibase 10.1103/PhysRevD.80.036009} {\bibfield  {journal}
  {\bibinfo  {journal} {\prd}\ }\textbf {\bibinfo {volume} {80}},\ \bibinfo
  {eid} {036009} (\bibinfo {year} {2009})},\ \Eprint
  {http://arxiv.org/abs/0906.1979} {arXiv:0906.1979 [hep-ph]} \BibitemShut
  {NoStop}%
\bibitem [{\citenamefont {{Chu}}\ \emph
  {et~al.}(2019{\natexlab{a}})\citenamefont {{Chu}}, \citenamefont
  {{Pradler}},\ and\ \citenamefont {{Semmelrock}}}]{Chu_19a}%
  \BibitemOpen
  \bibfield  {author} {\bibinfo {author} {\bibfnamefont {X.}~\bibnamefont
  {{Chu}}}, \bibinfo {author} {\bibfnamefont {J.}~\bibnamefont {{Pradler}}}, \
  and\ \bibinfo {author} {\bibfnamefont {L.}~\bibnamefont {{Semmelrock}}},\
  }\href {\doibase 10.1103/PhysRevD.99.015040} {\bibfield  {journal} {\bibinfo
  {journal} {\prd}\ }\textbf {\bibinfo {volume} {99}},\ \bibinfo {eid} {015040}
  (\bibinfo {year} {2019}{\natexlab{a}})}\BibitemShut {NoStop}%
\bibitem [{\citenamefont {{Chu}}\ \emph
  {et~al.}(2019{\natexlab{b}})\citenamefont {{Chu}}, \citenamefont {{Kuo}},
  \citenamefont {{Pradler}},\ and\ \citenamefont {{Semmelrock}}}]{Chu_19b}%
  \BibitemOpen
  \bibfield  {author} {\bibinfo {author} {\bibfnamefont {X.}~\bibnamefont
  {{Chu}}}, \bibinfo {author} {\bibfnamefont {J.-L.}\ \bibnamefont {{Kuo}}},
  \bibinfo {author} {\bibfnamefont {J.}~\bibnamefont {{Pradler}}}, \ and\
  \bibinfo {author} {\bibfnamefont {L.}~\bibnamefont {{Semmelrock}}},\ }\href
  {\doibase 10.1103/PhysRevD.100.083002} {\bibfield  {journal} {\bibinfo
  {journal} {\prd}\ }\textbf {\bibinfo {volume} {100}},\ \bibinfo {eid}
  {083002} (\bibinfo {year} {2019}{\natexlab{b}})},\ \Eprint
  {http://arxiv.org/abs/1908.00553} {arXiv:1908.00553 [hep-ph]} \BibitemShut
  {NoStop}%
\bibitem [{\citenamefont {{Adams}}\ \emph {et~al.}(1998)\citenamefont
  {{Adams}}, \citenamefont {{Sarkar}},\ and\ \citenamefont
  {{Sciama}}}]{Adams_98}%
  \BibitemOpen
  \bibfield  {author} {\bibinfo {author} {\bibfnamefont {J.~A.}\ \bibnamefont
  {{Adams}}}, \bibinfo {author} {\bibfnamefont {S.}~\bibnamefont {{Sarkar}}}, \
  and\ \bibinfo {author} {\bibfnamefont {D.~W.}\ \bibnamefont {{Sciama}}},\
  }\href {\doibase 10.1046/j.1365-8711.1998.02017.x} {\bibfield  {journal}
  {\bibinfo  {journal} {\mnras}\ }\textbf {\bibinfo {volume} {301}},\ \bibinfo
  {pages} {210} (\bibinfo {year} {1998})},\ \Eprint
  {http://arxiv.org/abs/astro-ph/9805108} {arXiv:astro-ph/9805108 [astro-ph]}
  \BibitemShut {NoStop}%
\bibitem [{\citenamefont {{Chen}}\ and\ \citenamefont
  {{Kamionkowski}}(2004)}]{Chen_04}%
  \BibitemOpen
  \bibfield  {author} {\bibinfo {author} {\bibfnamefont {X.}~\bibnamefont
  {{Chen}}}\ and\ \bibinfo {author} {\bibfnamefont {M.}~\bibnamefont
  {{Kamionkowski}}},\ }\href {\doibase 10.1103/PhysRevD.70.043502} {\bibfield
  {journal} {\bibinfo  {journal} {\prd}\ }\textbf {\bibinfo {volume} {70}},\
  \bibinfo {eid} {043502} (\bibinfo {year} {2004})},\ \Eprint
  {http://arxiv.org/abs/astro-ph/0310473} {arXiv:astro-ph/0310473 [astro-ph]}
  \BibitemShut {NoStop}%
\bibitem [{\citenamefont {{Slatyer}}(2016)}]{Slatyer_16}%
  \BibitemOpen
  \bibfield  {author} {\bibinfo {author} {\bibfnamefont {T.~R.}\ \bibnamefont
  {{Slatyer}}},\ }\href {\doibase 10.1103/PhysRevD.93.023527} {\bibfield
  {journal} {\bibinfo  {journal} {\prd}\ }\textbf {\bibinfo {volume} {93}},\
  \bibinfo {eid} {023527} (\bibinfo {year} {2016})},\ \Eprint
  {http://arxiv.org/abs/1506.03811} {arXiv:1506.03811 [hep-ph]} \BibitemShut
  {NoStop}%
\bibitem [{\citenamefont {{Liu}}\ \emph {et~al.}(2016)\citenamefont {{Liu}},
  \citenamefont {{Slatyer}},\ and\ \citenamefont {{Zavala}}}]{Liu_16}%
  \BibitemOpen
  \bibfield  {author} {\bibinfo {author} {\bibfnamefont {H.}~\bibnamefont
  {{Liu}}}, \bibinfo {author} {\bibfnamefont {T.~R.}\ \bibnamefont
  {{Slatyer}}}, \ and\ \bibinfo {author} {\bibfnamefont {J.}~\bibnamefont
  {{Zavala}}},\ }\href {\doibase 10.1103/PhysRevD.94.063507} {\bibfield
  {journal} {\bibinfo  {journal} {\prd}\ }\textbf {\bibinfo {volume} {94}},\
  \bibinfo {eid} {063507} (\bibinfo {year} {2016})},\ \Eprint
  {http://arxiv.org/abs/1604.02457} {arXiv:1604.02457 [astro-ph.CO]}
  \BibitemShut {NoStop}%
\bibitem [{\citenamefont {{Mangano}}\ and\ \citenamefont
  {{Serpico}}(2011)}]{Mangano_11}%
  \BibitemOpen
  \bibfield  {author} {\bibinfo {author} {\bibfnamefont {G.}~\bibnamefont
  {{Mangano}}}\ and\ \bibinfo {author} {\bibfnamefont {P.~D.}\ \bibnamefont
  {{Serpico}}},\ }\href {\doibase 10.1016/j.physletb.2011.05.075} {\bibfield
  {journal} {\bibinfo  {journal} {Physics Letters B}\ }\textbf {\bibinfo
  {volume} {701}},\ \bibinfo {pages} {296} (\bibinfo {year} {2011})},\ \Eprint
  {http://arxiv.org/abs/1103.1261} {arXiv:1103.1261 [astro-ph.CO]} \BibitemShut
  {NoStop}%
\bibitem [{\citenamefont {{Sabti}}\ \emph {et~al.}(2020)\citenamefont
  {{Sabti}}, \citenamefont {{Alvey}}, \citenamefont {{Escudero}}, \citenamefont
  {{Fairbairn}},\ and\ \citenamefont {{Blas}}}]{Sabti_20}%
  \BibitemOpen
  \bibfield  {author} {\bibinfo {author} {\bibfnamefont {N.}~\bibnamefont
  {{Sabti}}}, \bibinfo {author} {\bibfnamefont {J.}~\bibnamefont {{Alvey}}},
  \bibinfo {author} {\bibfnamefont {M.}~\bibnamefont {{Escudero}}}, \bibinfo
  {author} {\bibfnamefont {M.}~\bibnamefont {{Fairbairn}}}, \ and\ \bibinfo
  {author} {\bibfnamefont {D.}~\bibnamefont {{Blas}}},\ }\href {\doibase
  10.1088/1475-7516/2020/01/004} {\bibfield  {journal} {\bibinfo  {journal}
  {\jcap}\ }\textbf {\bibinfo {volume} {2020}},\ \bibinfo {eid} {004} (\bibinfo
  {year} {2020})},\ \Eprint {http://arxiv.org/abs/1910.01649} {arXiv:1910.01649
  [hep-ph]} \BibitemShut {NoStop}%
\bibitem [{\citenamefont {{Chu}}\ \emph {et~al.}(2020)\citenamefont {{Chu}},
  \citenamefont {{Kuo}},\ and\ \citenamefont {{Pradler}}}]{Chu_20}%
  \BibitemOpen
  \bibfield  {author} {\bibinfo {author} {\bibfnamefont {X.}~\bibnamefont
  {{Chu}}}, \bibinfo {author} {\bibfnamefont {J.-L.}\ \bibnamefont {{Kuo}}}, \
  and\ \bibinfo {author} {\bibfnamefont {J.}~\bibnamefont {{Pradler}}},\ }\href
  {\doibase 10.1103/PhysRevD.101.075035} {\bibfield  {journal} {\bibinfo
  {journal} {\prd}\ }\textbf {\bibinfo {volume} {101}},\ \bibinfo {eid}
  {075035} (\bibinfo {year} {2020})},\ \Eprint
  {http://arxiv.org/abs/2001.06042} {arXiv:2001.06042 [hep-ph]} \BibitemShut
  {NoStop}%
\bibitem [{\citenamefont {{Marocco}}\ and\ \citenamefont
  {{Sarkar}}(2020)}]{Marocco_20}%
  \BibitemOpen
  \bibfield  {author} {\bibinfo {author} {\bibfnamefont {G.}~\bibnamefont
  {{Marocco}}}\ and\ \bibinfo {author} {\bibfnamefont {S.}~\bibnamefont
  {{Sarkar}}},\ }\href@noop {} {\bibfield  {journal} {\bibinfo  {journal}
  {arXiv e-prints}\ ,\ \bibinfo {eid} {arXiv:2011.08153}} (\bibinfo {year}
  {2020})},\ \Eprint {http://arxiv.org/abs/2011.08153} {arXiv:2011.08153
  [hep-ph]} \BibitemShut {NoStop}%
\bibitem [{\citenamefont {{Chluba}}(2013{\natexlab{b}})}]{Chluba_13b}%
  \BibitemOpen
  \bibfield  {author} {\bibinfo {author} {\bibfnamefont {J.}~\bibnamefont
  {{Chluba}}},\ }\href {\doibase 10.1093/mnras/stt1733} {\bibfield  {journal}
  {\bibinfo  {journal} {\mnras}\ }\textbf {\bibinfo {volume} {436}},\ \bibinfo
  {pages} {2232} (\bibinfo {year} {2013}{\natexlab{b}})},\ \Eprint
  {http://arxiv.org/abs/1304.6121} {arXiv:1304.6121 [astro-ph.CO]} \BibitemShut
  {NoStop}%
\bibitem [{\citenamefont {{Tseliakhovich}}\ and\ \citenamefont
  {{Hirata}}(2010)}]{Tseliakhovich_10}%
  \BibitemOpen
  \bibfield  {author} {\bibinfo {author} {\bibfnamefont {D.}~\bibnamefont
  {{Tseliakhovich}}}\ and\ \bibinfo {author} {\bibfnamefont {C.}~\bibnamefont
  {{Hirata}}},\ }\href {\doibase 10.1103/PhysRevD.82.083520} {\bibfield
  {journal} {\bibinfo  {journal} {\prd}\ }\textbf {\bibinfo {volume} {82}},\
  \bibinfo {eid} {083520} (\bibinfo {year} {2010})},\ \Eprint
  {http://arxiv.org/abs/1005.2416} {arXiv:1005.2416 [astro-ph.CO]} \BibitemShut
  {NoStop}%
\bibitem [{\citenamefont {{Slatyer}}\ and\ \citenamefont
  {{Wu}}(2018)}]{Slatyer_18}%
  \BibitemOpen
  \bibfield  {author} {\bibinfo {author} {\bibfnamefont {T.~R.}\ \bibnamefont
  {{Slatyer}}}\ and\ \bibinfo {author} {\bibfnamefont {C.-L.}\ \bibnamefont
  {{Wu}}},\ }\href@noop {} {\bibfield  {journal} {\bibinfo  {journal} {ArXiv
  e-prints}\ } (\bibinfo {year} {2018})},\ \Eprint
  {http://arxiv.org/abs/1803.09734} {arXiv:1803.09734} \BibitemShut {NoStop}%
\end{thebibliography}%

\end{document}